\documentclass[%
 reprint,
 amsmath,amssymb,
 aps,
 pra,
floatfix,
]{revtex4-2}

\usepackage{graphicx}% Include figure files
\usepackage{dcolumn}% Align table columns on decimal point
\usepackage{bm}% bold math
\begin{document}

\preprint{APS/123-QED}

\title{Visualizing intense laser field driven electron dynamics in a multielectron molecule: dynamic electron localization, bonding properties and multiple ionization bursts.}

\author{L. Bauerle}
 \affiliation{JILA and Department of Chemistry, University of Colorado, Boulder, USA}
\author{A. Jaron}
\affiliation{
J. R. Macdonald Laboratory, Physics Department, Kansas State University,
Manhattan, USA}
\affiliation{JILA and Department of Physics, University of Colorado, Boulder, USA}

\date{\today}% It is always \today, today,

\begin{abstract}

The paper presents results and discussion of electron dynamics in molecules in which intense ultrafast pulse drives multiphoton ionization, high harmonic generation, and charge resonance due to the Rabbi flopping in the regime of Charge Resonance Enhanced Ionization (CREI) for the ground state molecular nitrogen cation. 

We show how ionization rates oscillate with oscillations of envelope reflecting modulation of the population of the excited molecular orbital. We observe the increase in ionization due to mechanisms analogous to Charge Resonance Enhanced Ionization (CREI) and also uncover also the time periods of suppressed ionization.  Electronic flux calculations illustrate these changes and help reveal multiple ionization bursts which are for the first time reported 
for a multielectron non-stretched/ non-dissociating molecule. 

Furthermore, we discuss the effects of ultrafast intense laser pulses on the bonding properties of a nitrogen molecular ion in these conditions through detailed analysis of electron density and electron density differences, time-dependent average local energy,  and time-dependent electron localization function. We illustrate the main characteristics of changes of the electron density that leads to laser induced dynamics of bonding and how the standard criteria lead to the transient properties of  double, triple, and lone pair bonding characteristics.   

%\begin{description}
%Secondary publications and information %retrieval purposes.
%You may use the \texttt{description} %environment to structure your abstract;
%use the optional argument of the \verb+\item+ %command to give the category of each item. 
%\end{description}
\end{abstract}

\maketitle

\section{Introduction}

Progress in controlling the properties of laser pulses is driving the efforts in direct studies of photochemical reactions in order to gain access to information about the dynamical properties and explore possibilities for ultrafast coherent control.  Often the properties of the laser pulses used in the experiments require the development of novel theoretical approaches that reflect the nature of the interaction. For example, the broad bandwidths and the ultrashort laser pulses require consideration of the molecular wavepacket propagation over multiple excited states \cite{Calegari_2016,Lepine_2013,Chini_2022,Peng_2019}. 
An important advantage of ultrashort pulses with a duration of up to few femtoseconds is that they offer access to the information and control of the chemical reactions at the timescales of electron dynamics. 
Ultrashort pulses mean that one can consider the observations as time-resolved snapshots of molecular wavepackets allowing for direct exploration of charge migration and charge localization \cite{He_2022}. 

The presence of the laser field in addition to the possibility of ionization, introduces a dressing that can be visualized classically as oscillations of the electrons within the system, synchronized to the oscillations of the electric field.  This simplified classical picture is often too simple to provide an adequate description of multielectron systems and one needs to consider dynamic quantum polarization and quantum excitations and their interplay in order to derive temporal properties for multielectron systems.  If we consider the set of field-free stationary states, the presence of the electric field of the laser pulse can cause significant mixing of the bound and continuum states during the interaction. This feature of the electron dynamics in the case of spatiotemporal grid simulations means a large spread of the wavepacket and strong modifications due to the interaction with the laser field as well as non-negligible ionization, which often requires complex absorbing potential applied at the edge of the grid.

Even without a significant probability of quantum excitation, electron dynamics can often reveal complex mechanisms due to the interference of multiorbital effects, electron polarization, excitation, and ionization. On the other hand even small population transfer to excited states can lead to the creation of the wavepacket for superposition states, which can result in novel interferences \cite{Xia_2016,Xia_2016_opt}.

Such superposition states commonly occur if one considers nonperturbative processes such as high-order harmonic generation and multiphoton ionization or if one considers experimental setups with multiple, often time-delayed, pulses such as coherent control schemes and experiments studying time delays \cite{Calegari_2016,EMO2018, Cattaneo_2016}.

Here we are interested in an example of dynamics that is not considered adiabatic and which can be observed using ultrashort pulses. We focus our studies on the effect of dynamic electron localization and investigate how one can visualize it in the multielectron system and what other consequences are related with such transient dynamic electron localization. 
The theoretical methods -- either basis set or spatio-temporal grid-based methods -- often consider dynamics with respect to initial and final field-free molecular states. Studies and visualization of what happens during the interaction with the intense laser pulse are often complicated by the fact that the dynamics cannot be understood only following populations in such states. Real and virtual processes involving bound and continuum states can lead to complicated patterns in the population of states during the interaction with the laser fields of nonperturbative intensities.  Here we consider  how populations can help understand the electron dynamics. 

Moreover we  study if the bonding properties are affected by the laser-induced transition where we remove an electron from the  inner valence antibonding orbital and excite it into bonding orbital of higher energy.

Ultrashort pulses due to the broad bandwidth often lead to excitations which result effectively in the superposition state being propagated during the interaction with the laser pulse, as such it has been considered a case in experiments on high harmonic spectroscopy of the hole dynamics \cite{Pluhar_2018, Li_2016}. Often theoretical investigations assume that the stage of excitation and/or ionization can be separated from the ''probing stage" for example via the process of HHG, even when it happens during the interaction with the same pulse. Such approximation assumes a single active electron and the rest of the electrons in the systems are treated as frozen ''spectators". 
Our results show that such an approximation should be applied with caution.

Depending on the peak laser intensity, it is observed that the electron wavepacket can exhibit signatures typical for dynamic electron localization. In the present paper, localization is meant as the effects in electron dynamics that can be characterized as nonadiabatic, not following the electric field of the laser pulse.  For a wavepacket it means there are periods of time when the wavepacket lags behind the dynamics driven by the electric field and for a diatomic molecule for example it results in localization on one side of the molecule.  
The effects are related to the interplay of the intense laser pulse interaction with the molecule and the evolution of the dressed superposition state. 
For H$_2^+$ such considerations used the charge resonance description \cite{Bandrauk_1993} to study the effects known later as  Charge Resonance Enhanced Ionization (CREI) \cite{Bandrauk_1995} and dynamic localization  \cite{Miller_2016}.
Later numerous studies considered the one electron picture to analyze dynamics for multielectron systems such as I$_2$ \cite{Gibson2018,Gibson2017}. 

In the present paper we describe the theoretical studies where we go beyond the one-electron approximation for a multielectron dynamic open system. For the diatomic molecule at the equilibrium bond length we study in detail the effects of the multielectron multiorbital nature of dynamic electron localization. Moreover, we investigate the interplay of the different contributions to the total density and Time-Dependent Electron Localization Function (TDELF) and relate it to bonding properties, multiphoton ionization, and Time-Dependent Average Local Ionization Energy (TDALIE). We envisage the present research as a step toward analyzing polyatomic systems, where localization can lead to interesting dynamic spatial properties of the molecular wavepacket with time scales for localization controlled by the laser pulse peak intensity. 

 Let us also note that electron localization properties has been studied in the context of x-ray pulses and  due to the nature of the core states the spatial information gained from electrons can be utilized to directly follow the separate atoms in a molecule. Typically if experiments are not performed with x-ray pulses one does not have access to spatial information or control molecular wavepackets, due to their delocalized character. Our study is motivated by the ideas of the control of electrons via control of dynamic localization in a molecule with UV or XUV laser pulses.

In the next section, we present definitions and theoretical methods used in the calculations. 

\section{Theoretical Methods}
Calculations are performed using time-dependent density functional theory (TDDFT) implemented in the open-source software Octopus \cite{Castro_2006}. 
Calculations proceed by the propagation of the molecular wavepacket and solving the set of time-dependent single electron Kohn-Sham equations for Kohn Sham orbitals $\phi_{i} $,
\begin{equation}
    \label{eq:tddft}
    \epsilon_{i} \phi_{i}(\textbf{r},t) = \left[ -\frac{1}{2} \nabla^{2} + V_{KS}(\textbf{r},t)  \right] \phi_{i}(\textbf{r},t).
\end{equation}
We have used the spin-polarized version of the TDDFT simulations for calculations for an open shell molecule, N$_2^+$. Moreover, because we needed to perform repeated calculations we have chosen the most computationally efficient method and used the local density approximation (LDA) for the exchange-correlation energy. 

The laser pulse is linearly polarized and the polarization is parallel to the molecular axis ($\hat{x}$-direction). The pulse envelope is assumed to be trapezoidal with a ramp up of 2 fs, a plateau of 8 fs, and a ramp down of 2 fs for a total pulse length of 12 fs. We assume that the nuclei are frozen during interaction with the laser pulse and the valence orbitals are active while the core orbitals are replaced by a pseudopotential \cite{Troullier_1991}.

In order to elucidate the effect of the laser pulse peak intensity on dynamic electron localization calculations were performed for peak intensities equal to $5 \times 10^{13}$ W/cm$^{2}$ and $2 \times 10^{14}$ W/cm$^{2}$). In order to study the effects of the resonance transition, simulations were performed assuming wavelentgh of  400 nm for 'on-resonance' case  and 600 nm for 'off-resonance' wavelengths. For the 600 nm wavelength pulse we keep the same pulse length of 12 fs. Consequently we have 9 o.c. for 400 nm and 6 o.c. for 600 nm for the lase pulse used in simulations. The peak intensity of the 600 nm pulse was $5 \times 10^{13}$ W/cm$^{2}$, matching one of the laser intensities of calculations performed for 400 nm wavelength pulse. The time-dependent Kohn-Sham equations are propagated using the Crank-Nicolson propagator \cite{Crank_1996} with a time-step of 0.02 a.u.. The code uses a grid-based method of TDDFT. In calculations, we used a parallel-piped grid that expands in the $\hat{x}$- and $\hat{y}$-direction from -40 to 40 a.u.. In the $\hat{z}$-direction, the grid extends from -30 to 30 a.u.. In all directions, the grid spacing is 0.3 a.u.. The molecular nitrogen cation was placed along the $\hat{x}$-axis with a bond length of 2.109 a.u., which was obtained from optimizing the geometry of the cation with Octopus. 
An absorbing potential of width equal to 5 a.u. is used and placed at the edge of the simulation box. In intense laser field simulations, even for the cases where there should be a minimal portion of the wavepacket making it to the edge of the simulation box, a complex absorbing potential is applied in order to decrease the interference effects from unphysical reflections from the boundaries and absorbing any part of the electronic wave-packet that makes it to the edge of the grid. This improves the precision of the properties calculated using the time-dependent wavepacket such as density difference, TDELF, TDALIE, and HHG. The portion of the wavepacket absorbed is considered when analyzing the total ionization.

The time-dependent dipole moment is calculated according to the equation
\begin{equation}
   \textbf{d}(t) = \int d\textbf{r} \textbf{r}\rho(\textbf{r},t).
\end{equation}
It is used to calculate the Fourier transform of time-dependent dipole moment, using a Blackman window and zero padding,
\begin{equation}
    \label{eq:FT}
    \hat{\textbf{d}}(\omega) = \int_{t_{1}}^{t_{2}} W(t) \textbf{d}(t) e^{-i \omega t} dt.
\end{equation}

Finally, the HHG spectrum is obtained from equation
\begin{equation}
    \label{eq:HHG}
    P(\omega) = \frac{\omega^4}{12 \pi \epsilon_0 c^3} \hat{\textbf{d}}(\omega) \hat{\textbf{d}}^{\star}(\omega).
\end{equation}

The primary dipole is along the laser polarization direction. For molecules it is possible to generate circularly or elliptically polarized HHG radiation for a linearly polarized driving laser pulse. In that case there is a significant time-dependent dipole in either of the perpendicular directions ($\hat{y}$ or $\hat{z}$). Ellipticity of harmonics generated from nitrogen molecule has been a subject of debate, since the Strong Field Approximation-based calculations predicted very small ellipticities for HHG from N$_2$ \cite{Levesque_2007}, while the experiments detected surprisingly different results \cite{Zhou_2009}. TDDFT simulation results for HHG ellipticities are in excellent agreement with experiments \cite{YX_thesis}. 

For systems with inversion symmetry, only odd harmonics are expected.  However, for systems where the same pulse that drives harmonic generation also drives the resonant transition each harmonic is accompanied by Mollow sidebands  \cite{Bandrauk_1995, Xia_2016_opt}. 
These sidebands are not integer multiples of the laser frequency, but a fractional of the fundamental frequency determined by the Rabi frequency.
For a large transition dipole,  it is expected that the intensity of the Mollow sidebands will be similar to that of the principal harmonics \cite{Xia_2016_opt}.

In order to observe the effects that the resonant transition preparing the superposition state has on the propagated molecular wavepacket, the density (spin-dependent and total) and TDELF (spin-dependent and total) are used. 
The time-dependent density can be used to derive properties such as the distribution of electrons in the molecule, in molecular and rescattering region, throughout the duration of the pulse and we anticipate to observe the effects of dynamic electron localization. TDELF and TDALIE are envisaged  to provide further quantitative information to understand  the modifications to molecular bonding and localization properties.

The electron localization function (ELF) is defined to provide a measure of the likelihood of finding an electron near a reference electron with the same spin \cite{Edgecombe_1990}. 
Traditionally ELF has been used for ground state systems to study atomic shells \cite{Edgecombe_1990} and bonding properties \cite{Larsson_2020, Silvi_2014}. 
ELF has also been used to analyze the con- and disrotary properties of photochemical reactions \cite{Guerra_2022} and  attosecond dynamics of irradiated molecules \cite{Parise_2018}.

In order to define ELF one starts with the probability of finding a same-spin electron near a reference electron at position $\textbf{r}$, which is given by formula
\begin{equation}
    \label{eq:gselocalization}
    D_{\sigma}(\textbf{r}) = \tau_{\sigma}(\textbf{r},t) - \frac{1}{4} \frac{[\nabla \rho_{\sigma}(\textbf{r},t)]^2}{\rho_{\sigma}(\textbf{r},t)^2},
\end{equation}
where $\tau_{\sigma}(\textbf{r})$ is the kinetic energy density described by single-particle orbitals $\varphi_{i \sigma}(\textbf{r})$,
\begin{equation}
    \label{eq:KEdensity}
    \tau_{\sigma}(\textbf{r}) = \sum_{i=1}^{N_{\sigma}} |\nabla \varphi_{i \sigma}(\textbf{r})|^2.
\end{equation}
This probability is then compared to the probability of finding a same-spin electron near a reference electron at position $\textbf{r}$ for a homogeneous electron gas of density $\rho_{\sigma}$,
\begin{equation}
    \label{eq:UEG}
    D_{\sigma}^0(\textbf{r}) = \frac{3}{5} (6 \pi)^{\frac{2}{3}} \rho_{\sigma}^{\frac{5}{3}}(\textbf{r}).
\end{equation}
In order to extend the ELF application to study dynamic properties of time-dependent wavefunctions one needs to modify the original $D_{\sigma}(\textbf{r})$ definition by the addition of a current density term to obtain 
\begin{equation}
    \label{eq:tdelocalization}
    D_{\sigma}(\textbf{r},t) = \tau_{\sigma}(\textbf{r},t) - \frac{1}{4} \frac{[\nabla \rho_{\sigma}(\textbf{r},t)]^2}{\rho_{\sigma}(\textbf{r},t)^2} - \frac{j_{\sigma}^2(\textbf{r},t)}{\rho_{\sigma}(\textbf{r},t)}
\end{equation}
where $j_{\sigma}(\textbf{r},t)$ is the absolute value of the current density defined as

\begin{equation}
   \frac{j_{\sigma}^{2}(\textbf{r},t)}{\rho_{\sigma}(\textbf{r},t)} = (\nabla \alpha_{\sigma}(\textbf{r},t))^{2} \rho_{\sigma}(\textbf{r},t)
    \label{eq:currentdensity}
\end{equation}
where $\alpha_{\sigma}(\textbf{r},t)$ is the phase of the complex single-particle wave function.
The current density term arises after assuming the molecular orbitals are complex-valued as shown in \cite{Gross_2005}. The  $\chi_{\sigma}(\textbf{r},t)$ is a dimensionless localization index that then compares the electron probability to that of the homogeneous electron gas by taking the ratio,
\begin{equation}
    \label{eq:ratio}
    \chi_{\sigma}(\textbf{r},t) = \frac{D_{\sigma}(\textbf{r},t)}{D_{\sigma}^0(\textbf{r},t)}
\end{equation}
The  ELF is then defined using formula 
\begin{equation}
    \label{eq:ELF}
    f_{ELF}(\textbf{r},t) = \frac{1}{1 + \chi_{\sigma}(\textbf{r},t)^2},
\end{equation}
in order to have values of $f_{ELF}(\textbf{r},t)$ ranging from [0,1]. If $f_{ELF}(\textbf{r},t)=0.5$ then the system corresponds to the homogeneous electron gas, and if $f_{ELF}(\textbf{r},t)=1$ then the system exhibits perfect localization.

The next section presents the results and discusses the detailed analysis of the results.

\section{Results}
\subsection{Projections}
We start presentation of the results  with the projections of the time-dependent wavefunctions onto the ground-state - initial wavefunction,
\begin{equation}
    \label{eq:projection}
    p = \int \phi_{i,\sigma}(\textbf{r},t)^{\star} \phi_{j,\sigma}(\textbf{r},t=0) d\textbf{r}.
\end{equation}
The projections illustrate the properties of the transition we driven with 400nm pulse and the corresponding ''superposition state". It also serves as a check of Rabi flopping in the context of multielectron wavefunction in the presence of the intense laser pulse. 
\begin{figure}[h!]
\includegraphics[width=1\columnwidth]{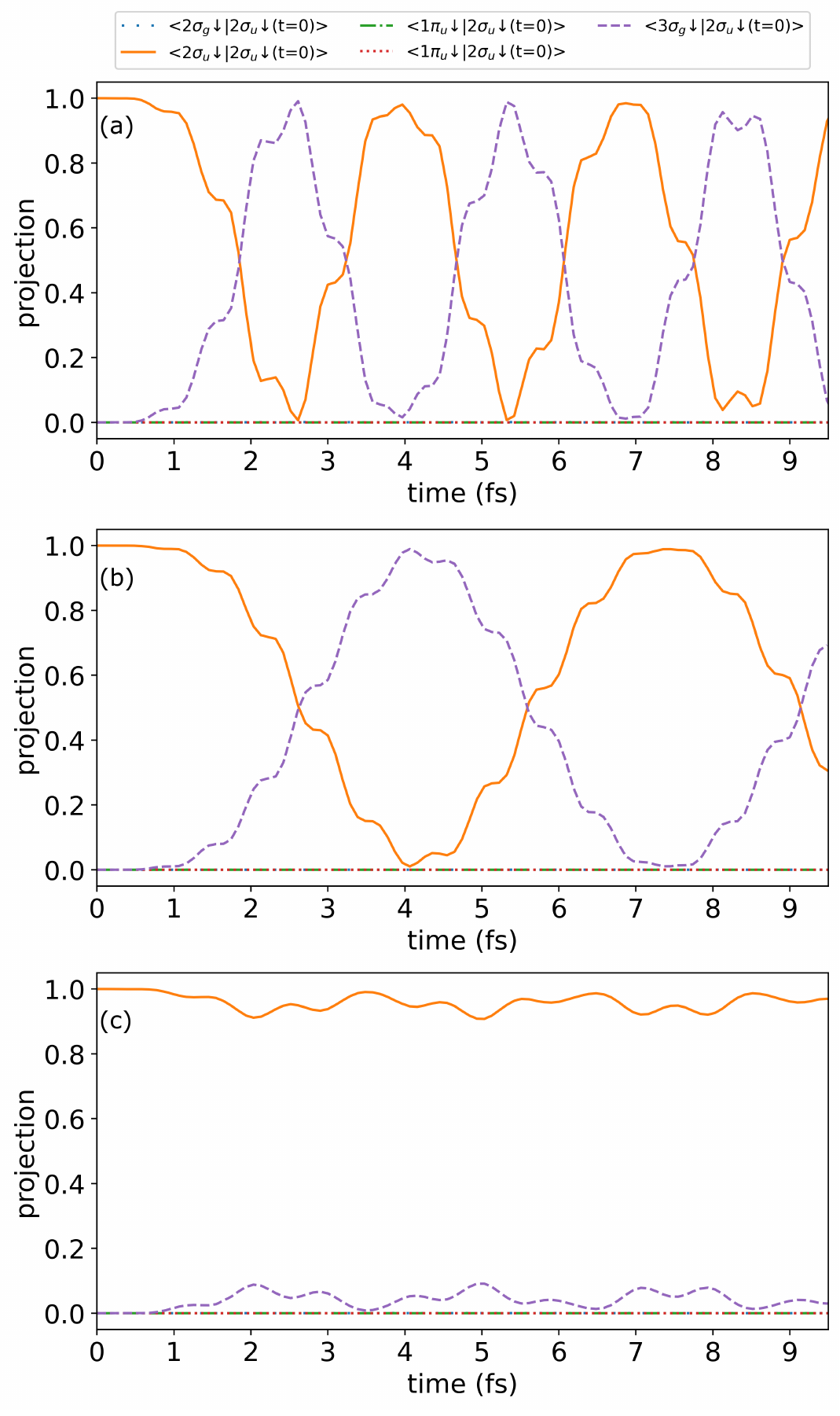}
\caption{\label{fig:proj} Time-dependent projections of the propagated wavefunction onto the ground state wavefunction of $2\sigma_{g,\downarrow}$ for (a) 400 nm, $2\times 10^{14} $ W/cm$^2$, (b) 400 nm, $5\times 10^{13} $ W/cm$^2$, and (c) 600 nm, $5\times 10^{13} $ W/cm$^2$.}
\end{figure}

\begin{figure}[h!]
\includegraphics[width=1\columnwidth]{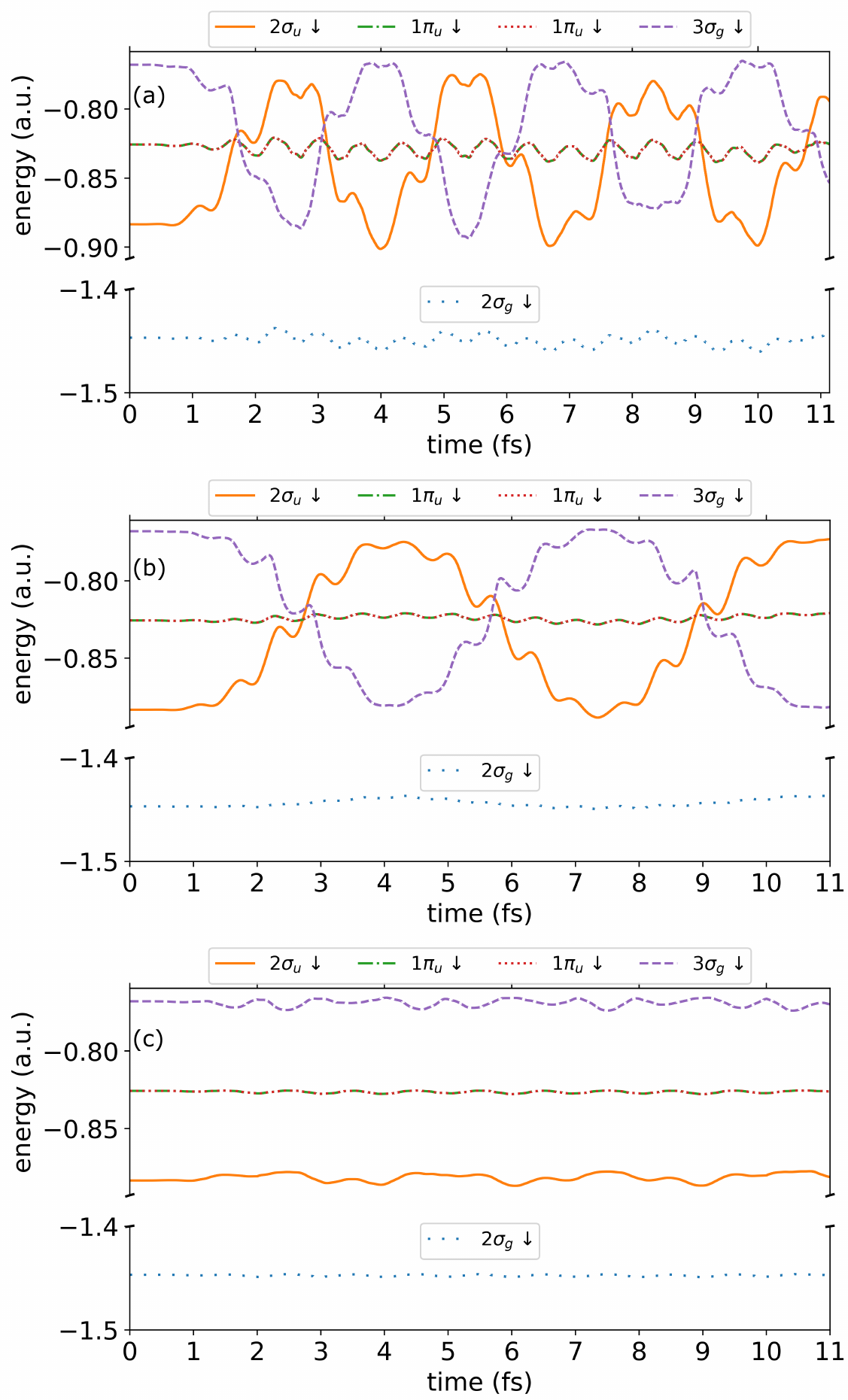}
\caption{\label{fig:energy} Time-dependent spin-down eigenvalues for the Kohn-Sham orbitals for (a) 400 nm, $2\times 10^{14}$ W/cm$^2$, (b) 400 nm, $5\times 10^{13}$ W/cm$^2$, and (c) 600 nm, $5\times 10^{13}$ W/cm$^2$.}    
\end{figure}

The projections are then compared to the time-dependent Kohn-Sham eigenvalues found using Equation \ref{eq:tddft} and shown in Figure \ref{fig:energy}.

When the superposition state is induced, the projections show that the $3\sigma_{g,\downarrow}(t)$ orbital has periods where it largely has $2\sigma_{u,\downarrow}$ character as seen in Fig. \ref{fig:proj} with the purple dashed line. This is an indication that the electron is transferred from the $2\sigma_{u,\downarrow}$ to $3\sigma_{g,\downarrow}$ orbital. Comparing the time-dependent projections in Fig. \ref{fig:proj} with the time-dependent eigenvalues in Fig. \ref{fig:energy} one can see that these periods of $3\sigma_{g,\downarrow}$ being the filled orbital coincide with the energy of that orbital decreasing and the energy of $2\sigma_{u,\downarrow}$ increasing.

For the ''uncoupled case" that is when one does not have a superposition state, the projections do not show the same mixing between $2\sigma_{u,\downarrow}$ and $3\sigma_{g,\downarrow}$, so the switching in the filled orbital for the 400 nm case is due to the field being coupled with the transition. An increase in peak intensity for the coupled case results in an increase in the frequency of this filled orbital switching. Throughout the rest of the paper, this change in eigenvalues will be used to illustrate the dynamics related to the transition.

\subsection{Electron Density}
\begin{figure}[h!]
\includegraphics[width=1\columnwidth]{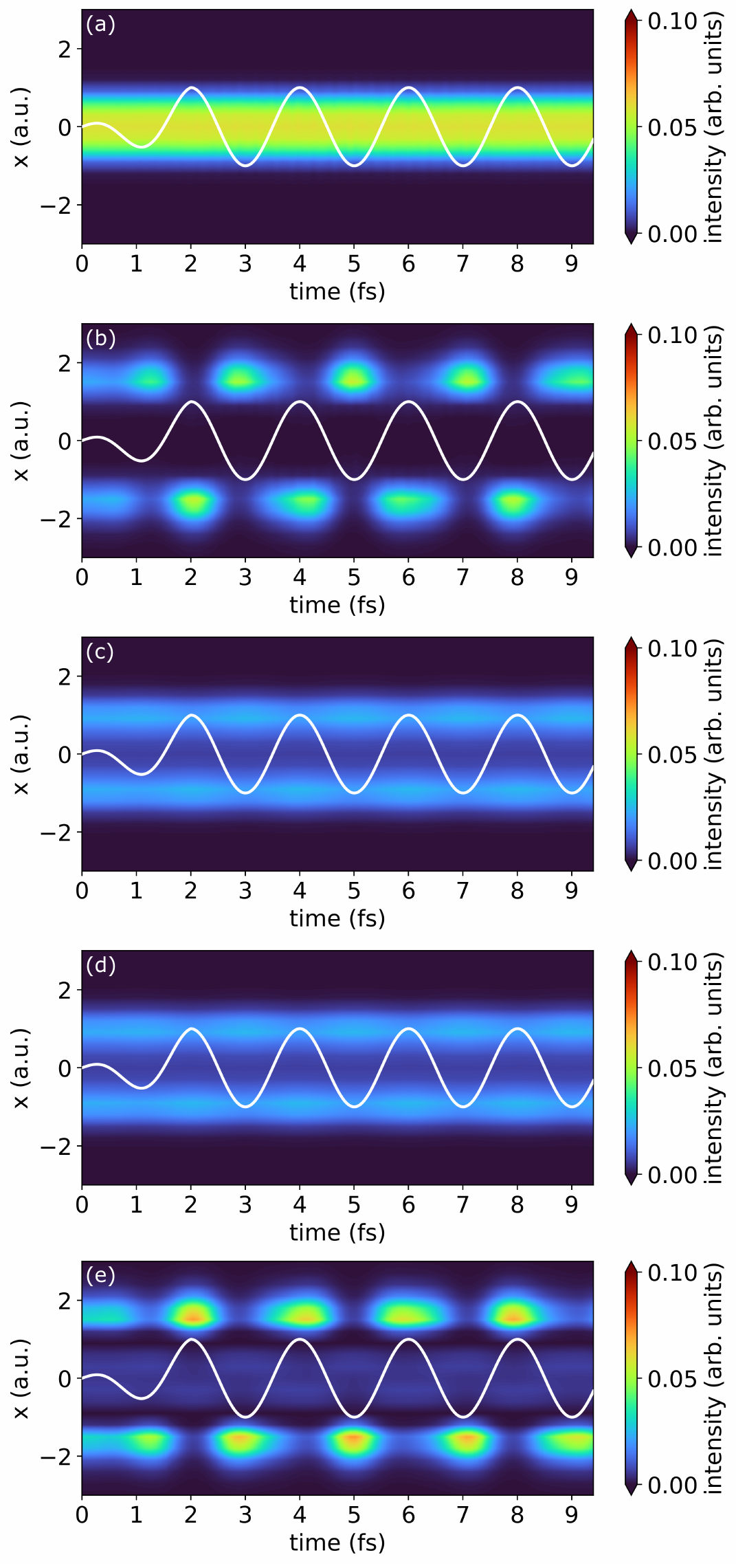}
\caption{\label{fig:600orb} Spin-down, time-dependent orbital densities for the off-resonance (600 nm) case with peak intensity $5 \times 10^{13} $ W/cm$^2$ in the polarization direction for (a) $2\sigma_g$, (b) $2\sigma_u$, (c) $1\pi_{u,a}$, (d) $1\pi_{u,b}$, and (e) $3\sigma_g$. The two other spatial dimensions are integrated over. Densities are plotted with the field in white to show time dependence.}   
\end{figure}

For the 600nm pulse, which is off-resonance with  transition, the orbital densities have maxima at the same time as the pulse as seen in Fig. \ref{fig:600orb}. This means that the integrated electron density for the $2\sigma_u$ is  showing the periodicity the same as the laser field as expected.

\begin{figure}[h!]
\includegraphics[width=1\columnwidth]{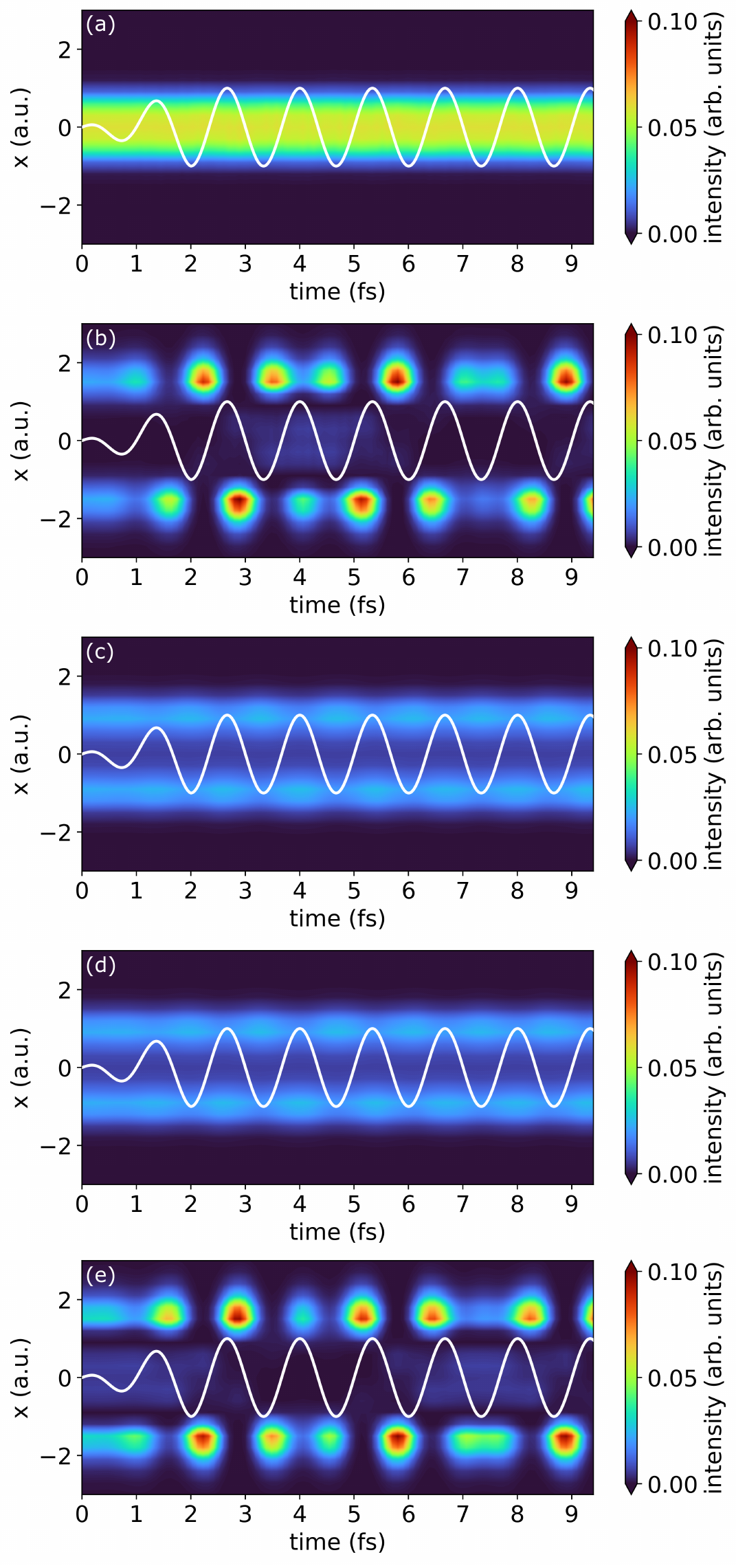}
\caption{\label{fig:400orb5e13}  
Spin-down, time-dependent orbital densities for the on-resonance (400 nm) case with peak intensity $5 \times 10^{13} W/cm^2$ in the polarization direction for (a) $2\sigma_g$, (b) $2\sigma_u$, (c) $1\pi_{u,a}$, (d) $1\pi_{u,b}$, and (e) $3\sigma_g$. The two other spatial dimensions are integrated over. Densities are plotted with the field in white to show time dependence.}   
\end{figure}

\begin{figure}[h!]
\includegraphics[width=1\columnwidth]{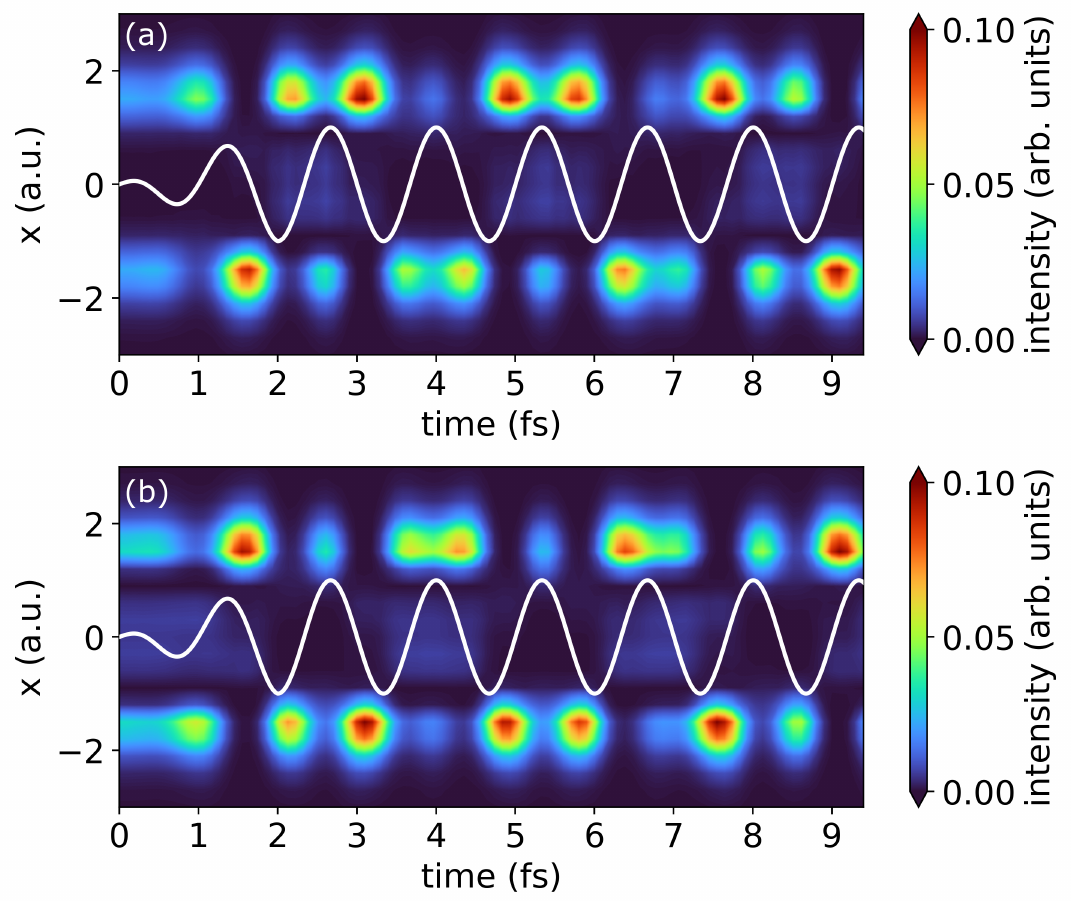}
\caption{\label{fig:400orb2e14} Spin-down, time-dependent orbital densities for the on-resonance (400 nm) case with peak intensity $2 \times 10^{14} $ W/cm$^2$ in the polarization direction for (a) $2\sigma_u$ and (b) $3\sigma_g$. The two other spatial dimensions are integrated over. Densities are plotted with the field in white to show time dependence.}   
\end{figure}

For the 400 nm pulse, which is on-resonance with the transition, the orbital densities have maxima occurring at times other than the peak intensities of the pulse. In Fig. \ref{fig:400orb5e13}(b), it can be seen that the peak of the pulse around 4 fs has not shifted the electron density as much as the cycle before it. While the field wants to push the electron density one way, the modulation due to the transition  leads to the localization of the electron cloud. By increasing the peak intensity, it can be seen in Fig. \ref{fig:400orb2e14}(b) that the instances of localization are now occurring more often than with the lower peak intensity case. 

\begin{figure}[h!]
\includegraphics[width=1\columnwidth]{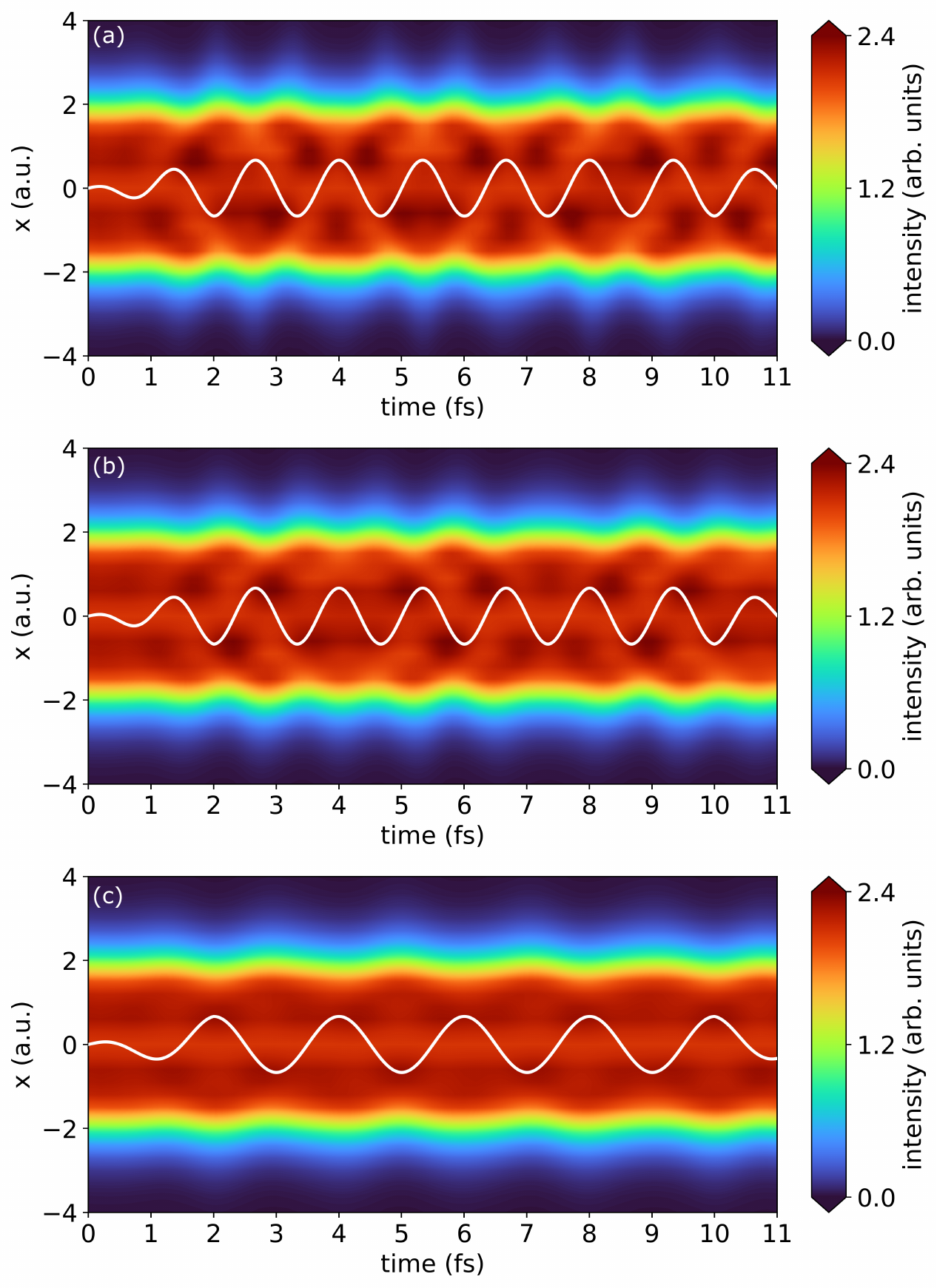}
\caption{\label{fig:TotalDensity} Total density dynamics in the polarization direction as a function of time for (a) 400 nm, $2\times 10^{14} $ W/cm$^2$, (b) 400 nm, $5\times 10^{13}$ W/cm$^2$, (c) 600 nm, $5\times 10^{13} $ W/cm$^2$. The two other spatial dimensions are integrated over.  The field is plotted in white to show time dependence.}  
\end{figure}

These instances of localization are still possible to be seen when looking at the total electron density. The total density is defined to be $\rho(\textbf{r},t) = \rho_{\uparrow}(\textbf{r},t) + \rho_{\downarrow}(\textbf{r},t)$. As shown in Fig. \ref{fig:TotalDensity}, there are instances of deeper red that match with the periods of localization seen at the orbital density level.

\begin{figure}[h!]
\includegraphics[width=1\columnwidth]{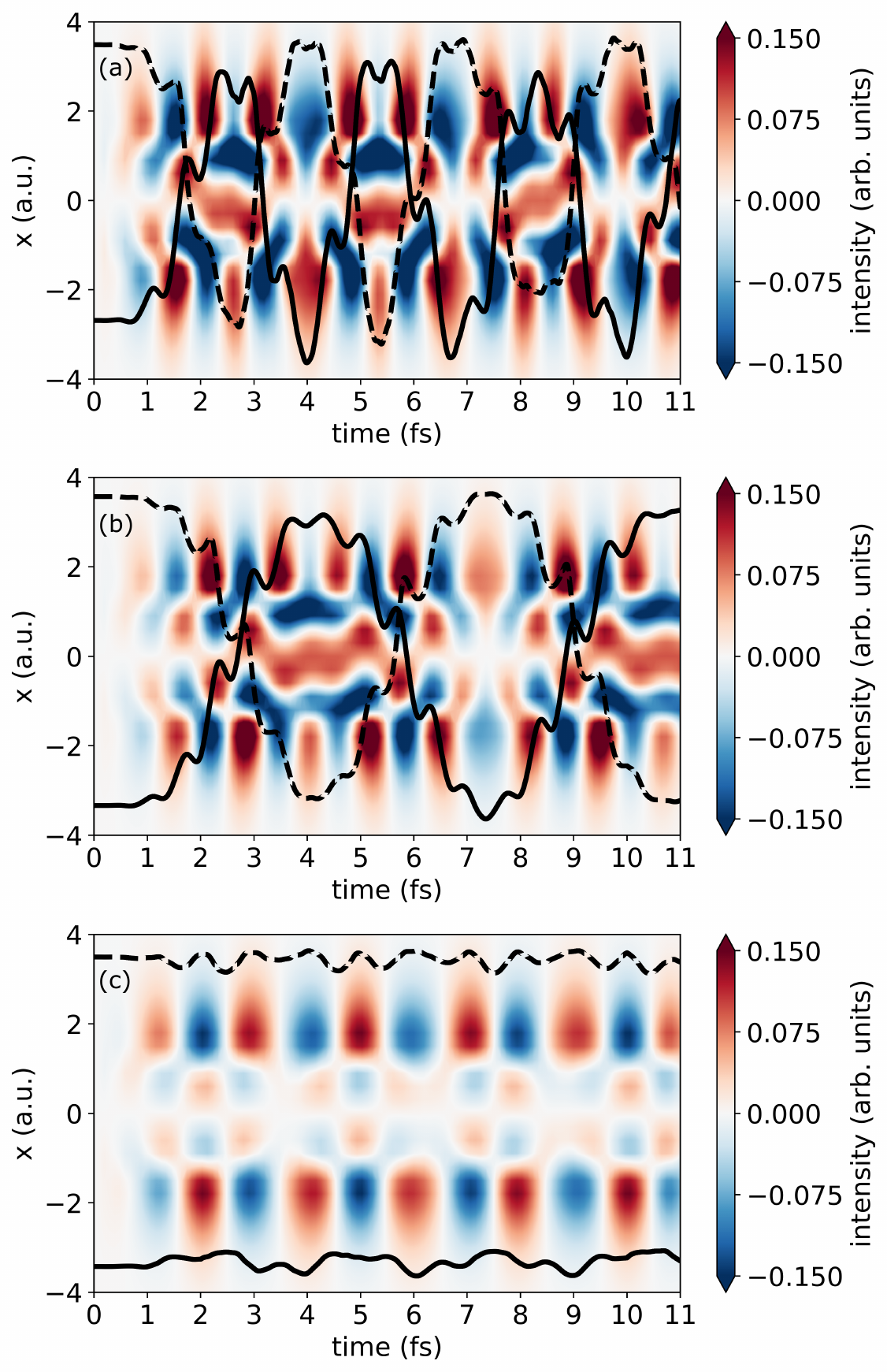}
\caption{\label{fig:TotalDensityDiff} Total density difference ($\rho(\textbf{r}, t) - \rho(\textbf{r}, t=0)$) in the polarization direction for (a) 400 nm, $2\times 10^{14} $ W/cm$^2$, (b) 400 nm, $5\times 10^{13} $ W/cm$^2$, (c) 600 nm, $5\times 10^{13} $ W/cm$^2$. The two other spatial dimensions were integrated out. The time-dependent eigenvalues for the $2\sigma_u$ (solid line) and $3\sigma_g$ (dashed line) orbitals are plotted as well.}
\end{figure}

In order to amplify these small changes seen in the total density, the density difference is used. The density difference is defined as $\rho_{diff}(\textbf{r},t) = \rho(\textbf{r},t)-\rho(\textbf{r},t=0)$. This is shown in Fig. \ref{fig:TotalDensityDiff} and it is now much easier to see the localization occurring since the range of values needed is much smaller. Periods of localization are now denoted by the red color not moving back and forth with the field. 

By revisiting the conditions for the orbital densities shown previously, it is shown that the switch to the total density difference picture still gives the same information about the localization, but now with an observable. To check that this localization process is occurring because of the transition, the time-dependent eigenvalues and projections were compared with the total density difference.
For example looking at the 600 nm case in Fig. \ref{fig:energy}(c), the lack of crossings means we do not expect any "orbital mixing" and is it off-resonance case, hence the electron density oscillates with the field periodicity  in Fig. \ref{fig:TotalDensity}(c).

The resonance case shows crossings between time dependent projections for orbitals occurring for both intensities considered in  the paper, as seen in Figs. \ref{fig:proj}(a) and \ref{fig:proj}(b). These quantities are reported in the Table \ref{tab:crossings}. This is also reflected in the eigenvalues as shown in Figs. \ref{fig:energy}(a) and \ref{fig:energy}(b) where $2\sigma_u$ and $3\sigma_g$ have periods of degeneracy. By comparing the time-dependent eigenvalues and projections with the density differences in Figs. \ref{fig:TotalDensityDiff}(a) and \ref{fig:TotalDensityDiff}(b), the periods of localization correspond to the filled orbital switching between $2\sigma_{u,\downarrow}$ and $3\sigma_{g,\downarrow}$.

\begin{table}[b]
\caption{\label{tab:crossings} Time instances of crossings of eigenvalues for $2\sigma_{u,\downarrow}$ and $3\sigma_{g,\downarrow}$ for a wavelength of 400 nm.}
\begin{ruledtabular}
\begin{tabular}{ccc}
    I ($ W/cm^2$)  & $2\sigma_{u,\downarrow}$ to $3\sigma_{g,\downarrow}$ (fs) &  $3\sigma_{g,\downarrow}$ to $2\sigma_{u,\downarrow}$ (fs) \\
\hline
    $5\times 10^{13}$ & 2.7791 & 5.7279  \\
                     & 8.9959 &         \\
    $2 \times 10^{14}$ & 1.8769 & 3.0815 \\ 
                     & 3.9306 & 4.3362 \\
                     & 5.5414 & 6.6785 \\
                     & 8.0331 & 11.1388
\end{tabular}
\end{ruledtabular}     
\end{table}

There is an increase in the number of crossings for the higher peak intensity case due to the increase in the Rabi frequency, which would correspond to the increase in localization events seen in the density and density difference. This increase in localization also allows for the ability to assign spatial features to the localization events and changes in orbital filling. For instance, as seen in Fig. \ref{fig:TotalDensityDiff}(a), at 4 fs when the $2\sigma_u$ is filled, the electron is localized on the top nitrogen atom and around 5.5 fs when $3\sigma_g$ is filled, the electron is localized on the bottom nitrogen atom. This is not the case for the lower peak intensity case, where both periods of electron localization correspond to the density focused on the bottom nitrogen as shown in Fig. \ref{fig:TotalDensityDiff}(b). This means that for certain intensities, periods of localization on the top or the bottom nitrogen atom correlate to the current state of the system.

\subsection{Electron Localization Function}
In addition to the density, the dynamical localization properties can be visualized using the time-dependent electron localization function (TDELF). 

\begin{figure}[h!]
\includegraphics[width=0.9\columnwidth]{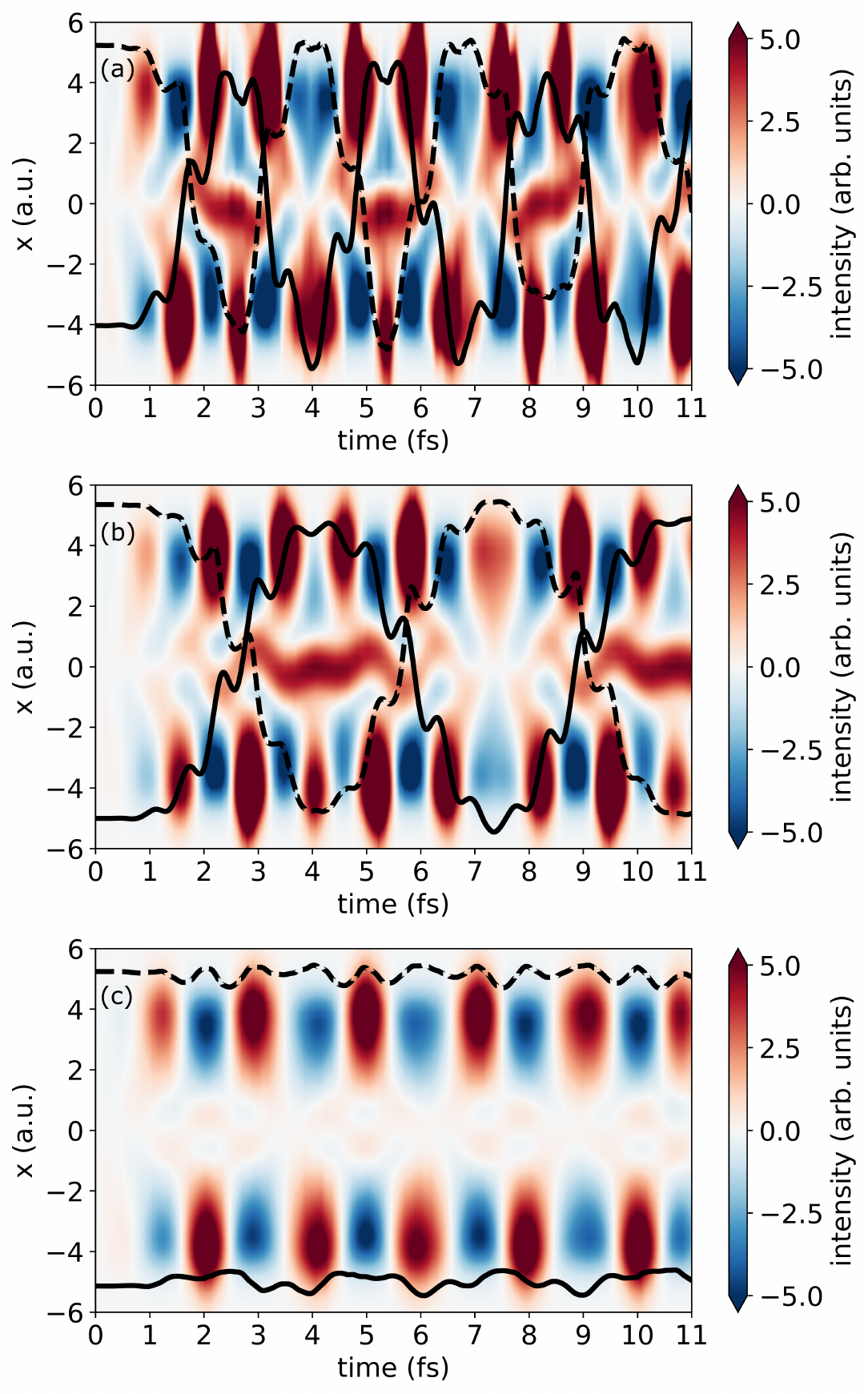}
\caption{\label{fig:TotalELFDiff} Total ELF ($f_{ELF}(\textbf{r},t) - f_{ELF}(\textbf{r}, t=0)$) in the polarization direction for (a) 400 nm, $2\times 10^{14} $ W/cm$^2$, (b) 400 nm, $5\times 10^{13} $ W/cm$^2$, (c) 600 nm, $5\times 10^{13} $ W/cm$^2$. The two other spatial dimensions are integrated out. The time-dependent eigenvalues for the $2\sigma_u$ (solid line) and $3\sigma_g$ (dashed line) orbitals are plotted as well.}
\end{figure}

By looking at the TDELF difference, it can be seen in Fig. \ref{fig:TotalELFDiff} that the changes in orbital occupation correspond to changes in TDELF. For instance, during periods where $3\sigma_g$ is occupied, there is an increase in the electron localization between the nitrogen atoms. This is the same dynamics that was seen with the total density difference plots, where there are periods of increased electron density in the bonding region. One can observe that visualization using TDELF emphasizes these changes and that bonding region is affected by the dynamic localization and additional modulations can be seen very clearly and one can see the modulation has the same time periodicity as  the eigenvalues and orbital projections. The effects on the lone pairs are also emphasized in addition to the bonding region because TDELF is able to describe lone pairs and radical character better than the electron density. For systems with lone pairs and/or radical electrons, such as N$_2^+$, using TDELF can be a better visualization of the interaction of non-bonding electrons in time.

\subsection{Dynamic localization effect on the bonding}

Dynamic electron localization changes the molecular potential locally and hence has an effect on the local bonding properties. This could have further consequences on the chemical reactions that could, for example, by modified when initiated during different stages of localization. Here we present results that illustrate how the bonding properties are affected by the ultrafast intense laser pulse and by the nonadiabatic changes due to dynamic electron localization. In particular, these changes can be related to the time dependence of the instantaneous orbital energies. 

Calculations were performed for different degrees of mixing between $2\sigma_{u,\downarrow}$ and $3\sigma_{g,\downarrow}$. In addition to the changes in projections and eigenvalues indicating the changes in orbital filling, there are characteristic features seen in both the density and ELF slices in the xy-plane. These characteristics can be used as a visual representation of the state of the system when expanding to time-dependent calculations.

A range of 0.6 to 1 for the values of ELF was used because this range encompasses the covalent and lone pair values of ELF. According to \cite{Larsson_2020}, for values of ELF greater than 0.7 there are localized electrons indicating bonding and the ELF values closest to 1 represent lone pair electrons. In \cite{Melin_2003}, the isosurfaces for radical organic systems were plotted for the ELF value of 0.6, indicating a lower value is needed when discussing radical systems. Melin and Fuentealba also noted in \cite{Melin_2003} that with spin-dependent ELF, high values were now able to represent a radical electron. 

\begin{figure}[h!]
\includegraphics[width=1\columnwidth]{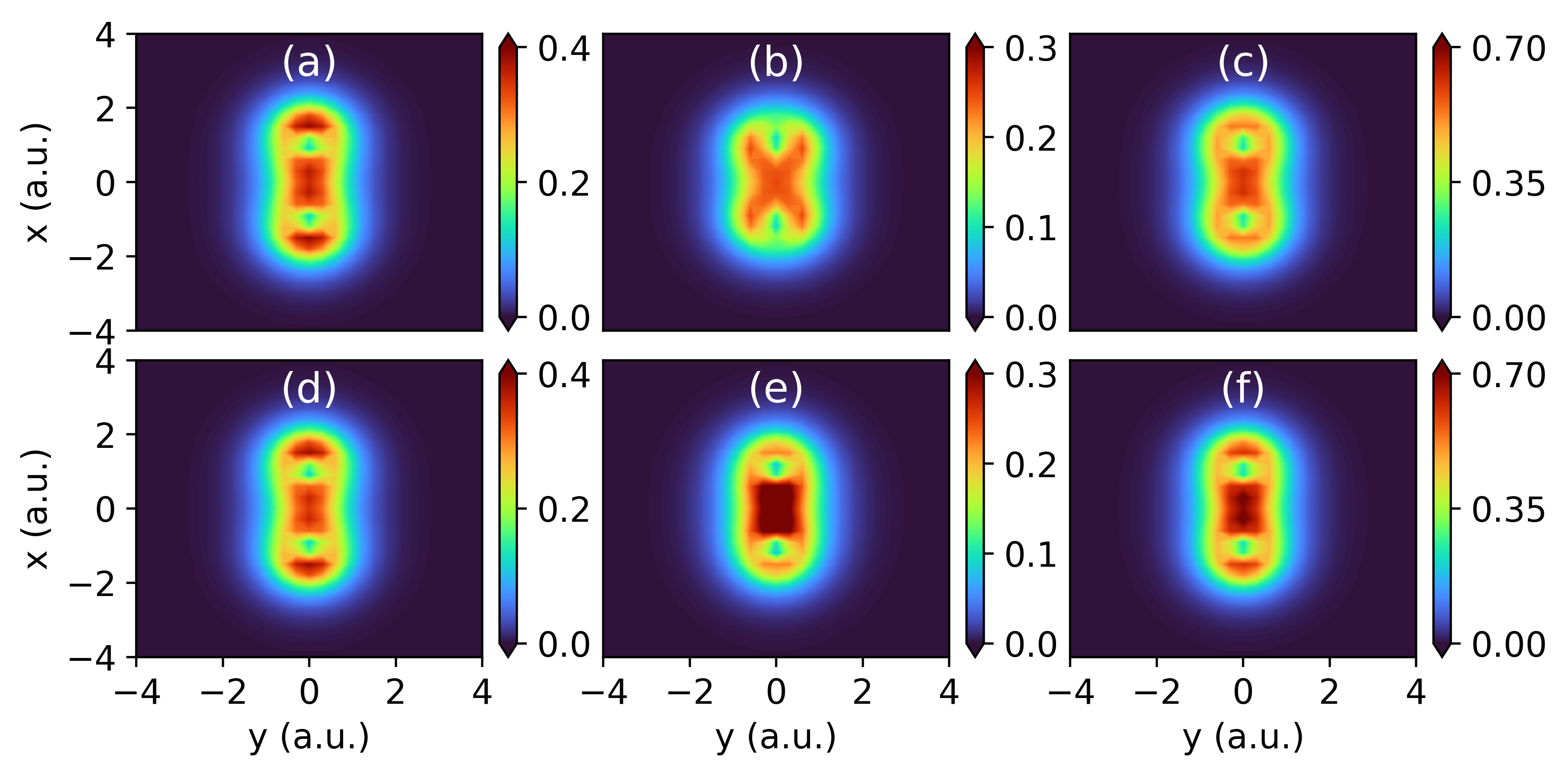}
\caption{\label{fig:mixing_density} Cross sections (z=0) of density where (a-c) are the spin-up (a), spin-down (b), and total (c) for the $2\sigma_{u,\downarrow}$ filled case and (d-f) are the spin-up (d), spin-down (e), and total (f) for the $3\sigma_{g,\downarrow}$ filled case.}
\end{figure}

\begin{figure}[h!]
\includegraphics[width=1\columnwidth]{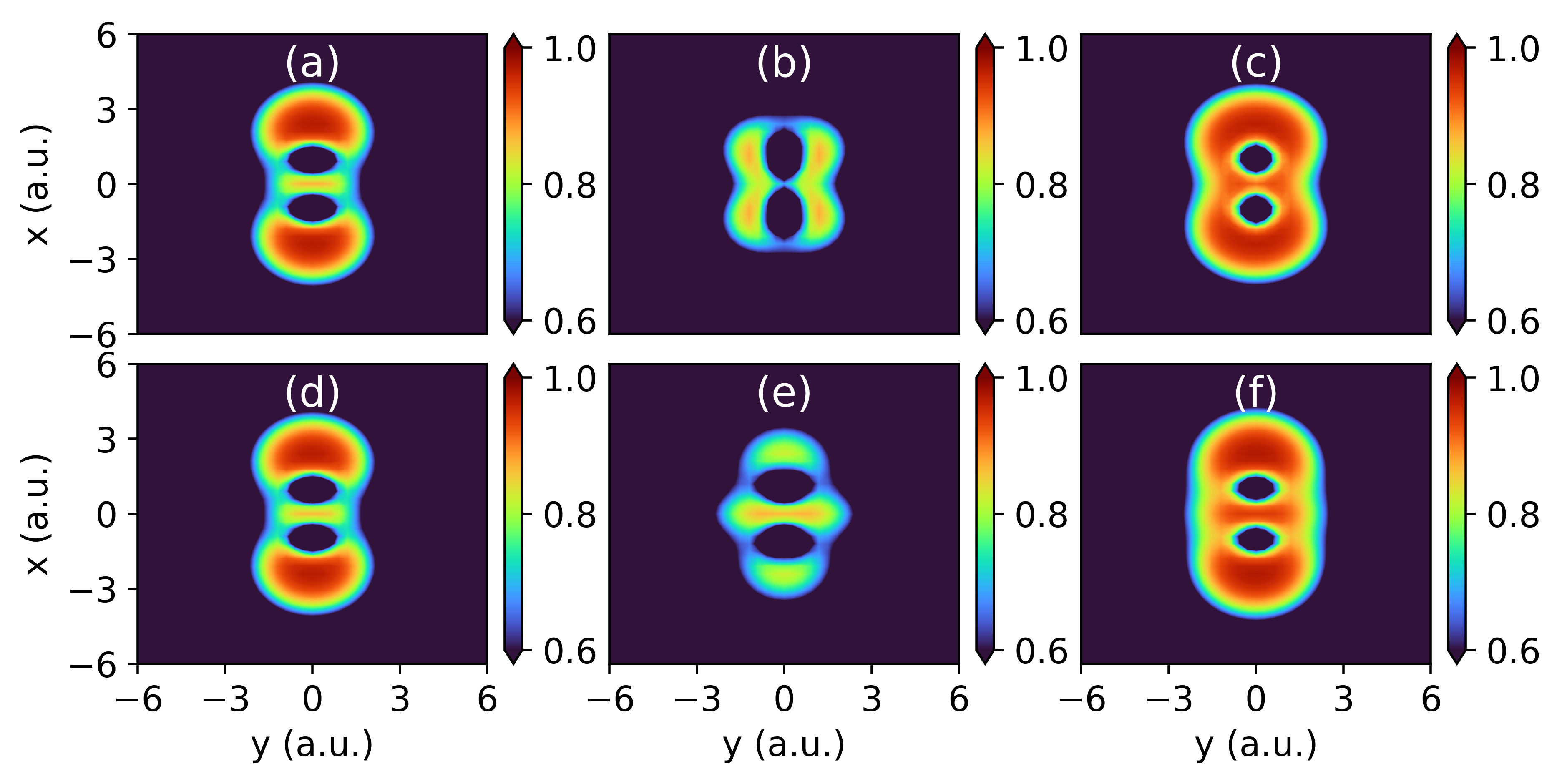}
\caption{\label{fig:mixing_elf} Cross sections (z=0) of ELF where (a-c) are the spin-up (a), spin-down (b), and total (c) for the $2\sigma_{u,\downarrow}$ filled case and (d-f) are the spin-up (d), spin-down (e), and total (f) for the $3\sigma_{g,\downarrow}$ filled case.}
\end{figure}

Cross sections of the ELF and density  in the xy-plane calculated in order to see how the bonding changes for N$_2^+$. According to \cite{Silvi_2014}, the shape of ELF corresponds to the bond order for a molecule. In three dimensions, a single bond has an almost spherical shape, a double bond has a peanut shape, and a triple bond has a torus shape. These shapes can also be distinguished in two dimensions as discussed in \cite{Larsson_2020} with respect to ethane, ethene, and ethyne. When looking at the slices along the xy-plane for these molecules, a single bond (ethane) has a rice grain shape, a double bond (ethene) has a tube shape, and a triple bond (ethyne) has a dumbbell shape.

For the ELF slices, there is a characteristic increase in the width of ELF between the two nitrogen atoms for both the total ELF and spin-down ELF as seen in Fig. \ref{fig:mixing_elf} as the system increases $3\sigma_{g, \downarrow}$ character. There is also a change in the shape of ELF in that bonding region. If $2\sigma_{u,\downarrow}$ is filled the bonding region of the total ELF has a dumbbell shape, whereas if $3\sigma_{g,\downarrow}$ is filled the bonding region has a tube shape. Both features could be used as a visual indication of $3\sigma_{g,\downarrow}$ being filled.

For the density slices, there is a characteristic increase in the density between the two nitrogen atoms for both the total density and spin-down density as seen in Fig. \ref{fig:mixing_density}. This could be used as an indication of $3\sigma_{g,\downarrow}$ being filled.

The trends seen for the static cases of ELF slices and electron density slices is used to better understand the changes seen during the interaction with the laser pulse. To have more instances of illustrative cases (changes to orbital filling and extrema of laser field), the on-resonance 5$\times$10$^{13}$ W/cm$^2$ case will be discussed further.

While $2\sigma_{u,\downarrow}$ is the filled orbital, the ELF slice shows a dumbbell shape indicating the bonding has triple bond character, as expected from the Lewis structure. For this case the density was integrated in the bonding region and this calculation gave a value of 5.9145 confirming the triple bond. While $3\sigma_{g,\downarrow}$ is the filled orbital, the ELF slices show a rice grain shape indicating the bond shows more double bond character. This shape is what is seen in the excited N$_2^+$ static calculation, confirming the filling of the orbitals. 

\begin{figure}[h!]
\includegraphics[width=0.9\columnwidth]{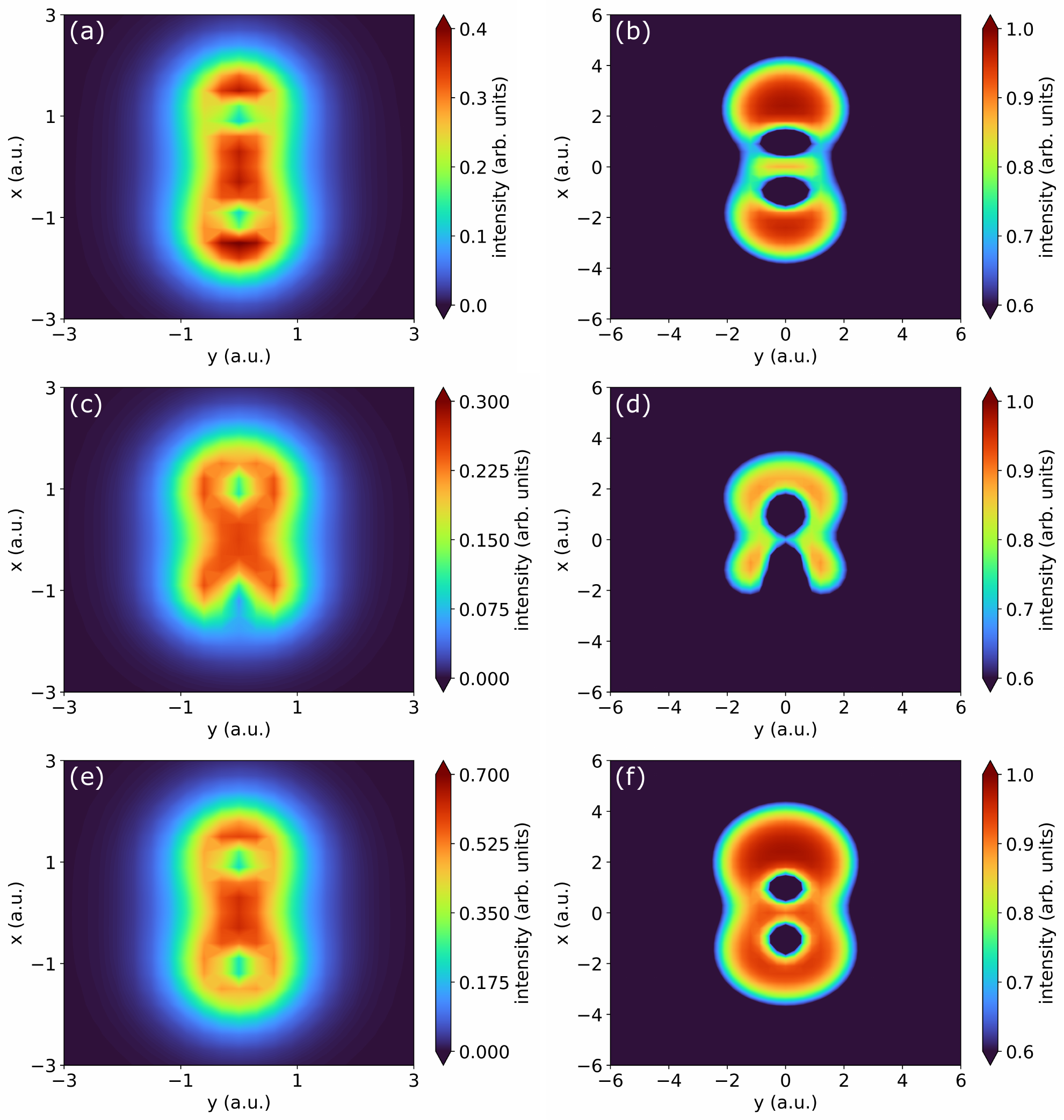}
\caption{\label{fig:iter2000_2} Cross sections (z=0) of (a) spin-up density, (c) spin-down density, (e) total density, (b) spin-up ELF, (d) spin-down ELF, and (f) total ELF along the xy-plane for the on-resonance, 5$\times$10$^{13}$ W/cm$^2$ case. These slices were taken at 0.9676 fs, which is near a zero of the field and just before there is significant mixing between  $2\sigma_u$ and $3\sigma_g$.}
\end{figure}

During the ramp-up of the pulse for the $5\times10^{13}$ W/cm$^2$ coupled case, there are hardly any differences in the density and ELF, likely due to the low peak intensity of the pulse during this time. Sloshing with the field can start to be seen around 1 fs as seen in Fig. \ref{fig:iter2000_2} as the peak intensity has steadily increased over the first optical cycle. It is around this time that the field amplitude greatly increases and soon after this time the mixing between $2\sigma_u$ and $3\sigma_g$ begins as shown in Fig. \ref{fig:proj} with the decrease in $2\sigma_{u,\downarrow}(t)$ having $2\sigma_{u,\downarrow}(t=0)$ character. While the field amplitude is still low around the first zero of the laser field, Figs. \ref{fig:iter2000_2}(a,c,e) show that the density is asymmetric, indicating the localization of the electron density. At this point $2\sigma_u$ is the filled orbital which can be seen in Fig. \ref{fig:iter2000_2}(f) by the tube shape of the total ELF. 

\begin{figure}[h!]
\includegraphics[width=0.9\columnwidth]{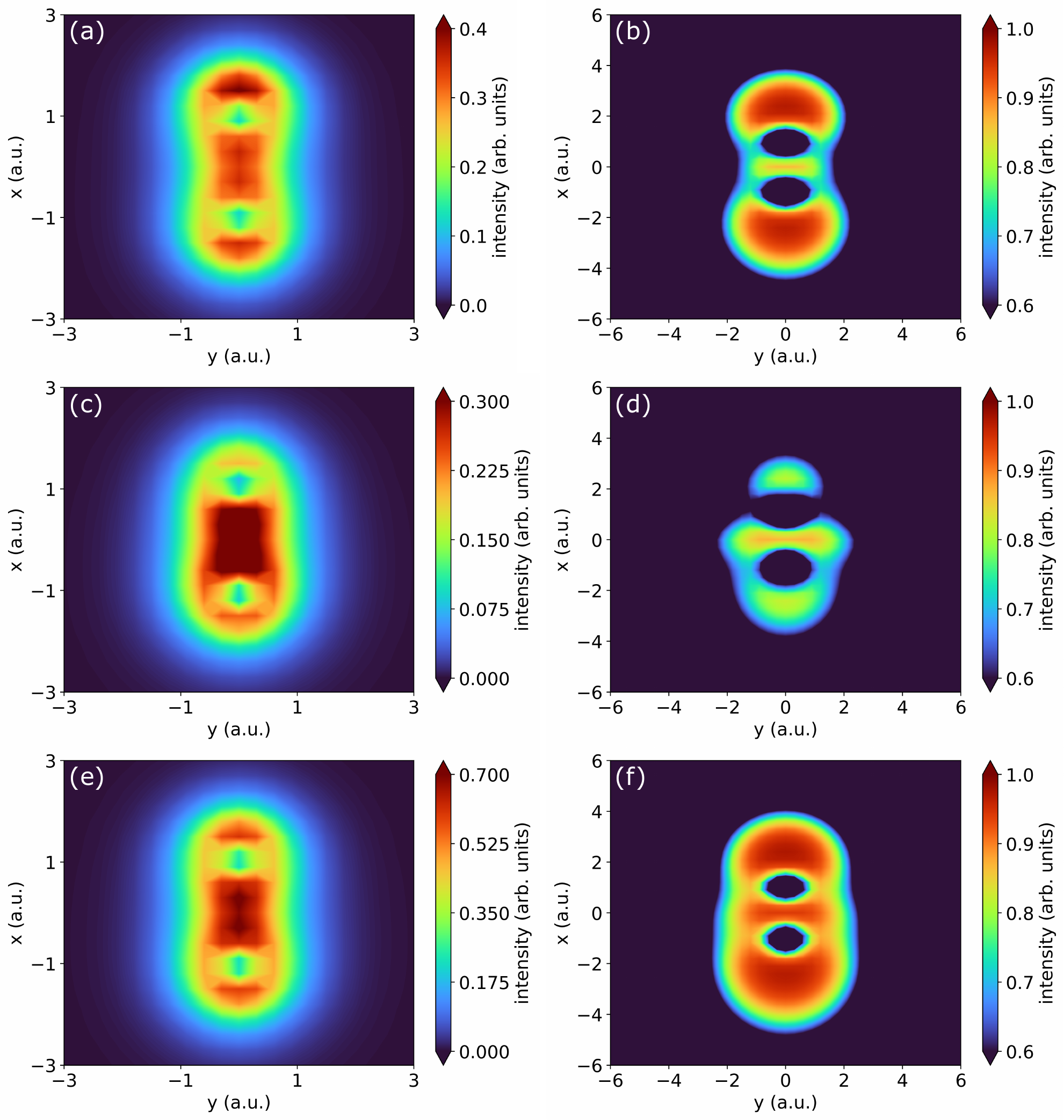}
\caption{\label{fig:iter5400_2} Cross sections (z=0) of (a) spin-up density, (c) spin-down density, (e) total density, (b) spin-up ELF, (d) spin-down ELF, and (f) total ELF along the xy-plane for the on-resonance, 5$\times$10$^{13}$ W/cm$^2$ case. These slices were taken at 2.6125 fs, which is near a crossing between $2\sigma_u$ and $3\sigma_g$ and a maxima of the field.}
\end{figure}

During the initial period of having $2\sigma_u$  filled there is little to no change in the ELF between the nitrogen atoms. This indicates that the density sloshing back and forth with the field is not having an effect on the bonding. As there is more mixing with the $3\sigma_g$ orbital, changes between the nitrogen atoms can be observed for both the density and the ELF. These changes can be see in Fig. \ref{fig:iter5400_2} which shows slices of ELF and density near the crossing where the orbital filling switches from $2\sigma_{u,\downarrow}$ to  $3\sigma_{g,\downarrow}$. There is an increase in the density between the atoms shown in Figs. \ref{fig:iter5400_2}(a,c,e) and the ELF begins to widen as seen in Figs. \ref{fig:iter5400_2}(b,d,f). This matches the trends seen for the excited N$_2^+$ static case discussed previously.

\begin{figure}[h!]
\includegraphics[width=0.9\columnwidth]{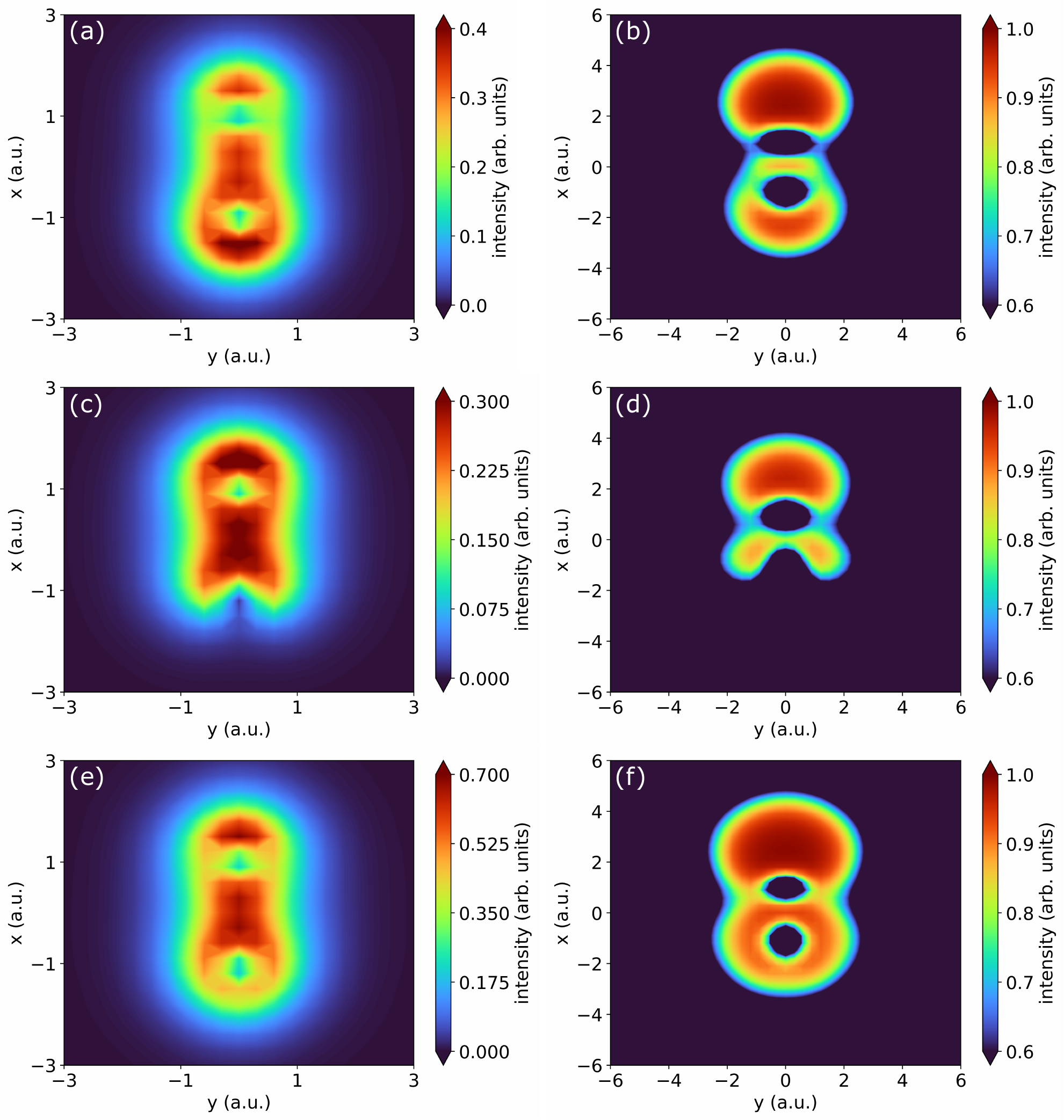}
\caption{\label{fig:iter6200_2} Cross sections (z=0) of (a) spin-up density, (c) spin-down density, (e) total density, (b) spin-up ELF, (d) spin-down ELF, and (f) total ELF along the xy-plane for the on-resonance, 5$\times$10$^{13}$ W/cm$^2$ case. These slices were taken at  2.9995 fs, which is near a zero of the field and the onset of the period in which $3\sigma_g$ is considered filled.}
\end{figure}

The increase in density between atoms can be seen around 2.9995 fs, which is approximately 1 fs after the crossing between states and at a zero of the field. This increase in density in the bonding region is shown in Fig. \ref{fig:iter6200_2}(e) and it can also be seen that there is localization as the electron density is largely on the top nitrogen atom. It is here that the features indicating $3\sigma_{g,\downarrow}$ being the filled orbital can be seen more clearly. Figure \ref{fig:iter6200_2}(f) shows that the bonding region in the total ELF slices widens and has more of a tube shape, indicative of the orbital filling and change in bonding properties of the molecule. This change in shape indicates that there is an increase in double bond character when the $3\sigma_{g,\downarrow}$ is the filled orbital.

\begin{figure}[h!]
\includegraphics[width=0.9\columnwidth]{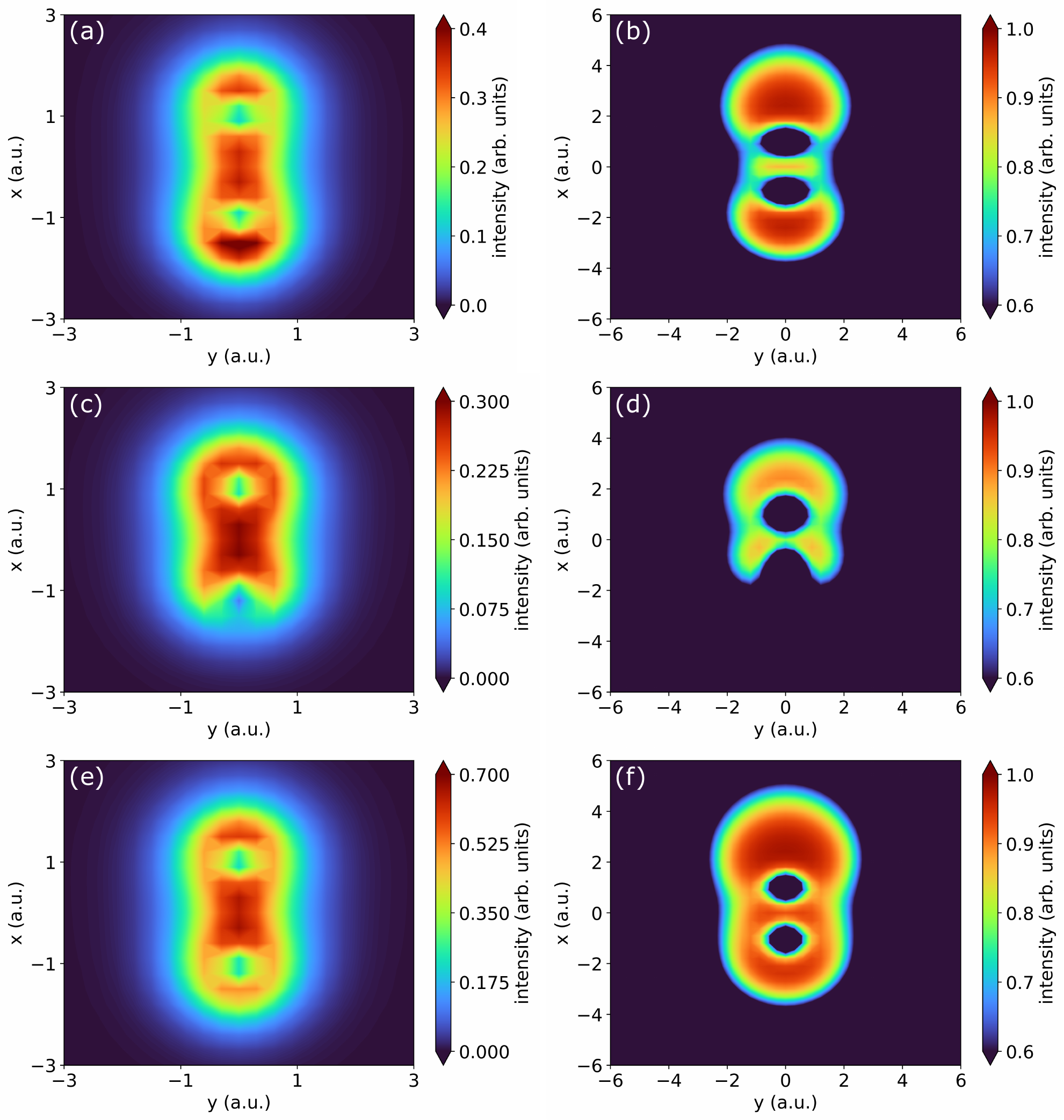}
\caption{\label{fig:iter6800_2} Cross sections (z=0) of (a) spin-up density, (c) spin-down density, (e) total density, (b) spin-up ELF, (d) spin-down ELF, and (f) total ELF along the xy-plane for the on-resonance, 5$\times$10$^{13}$ W/cm$^2$ case. These slices were taken at 3.2898 fs, which is near a minimum of the laser field and during the period when $3\sigma_g$ is considered filled. }
\end{figure}

Around 3.2898 fs, the total density difference indicates that there is a further increase in the density around the ''top" nitrogen atom and bonding region, while the ''bottom” nitrogen atom remains largely unchanged, as seen in Fig. \ref{fig:iter6800_2}(e). This trend of an increase in density on one side of the molecule with little to no change on the other side occurs throughout the $3\sigma_g$ filled region, another indication of localization. Near the minimum in the laser field the total ELF still has the tube shape shown in Fig. \ref{fig:iter6800_2}(f).

\begin{figure}[h!]
\includegraphics[width=0.9\columnwidth]{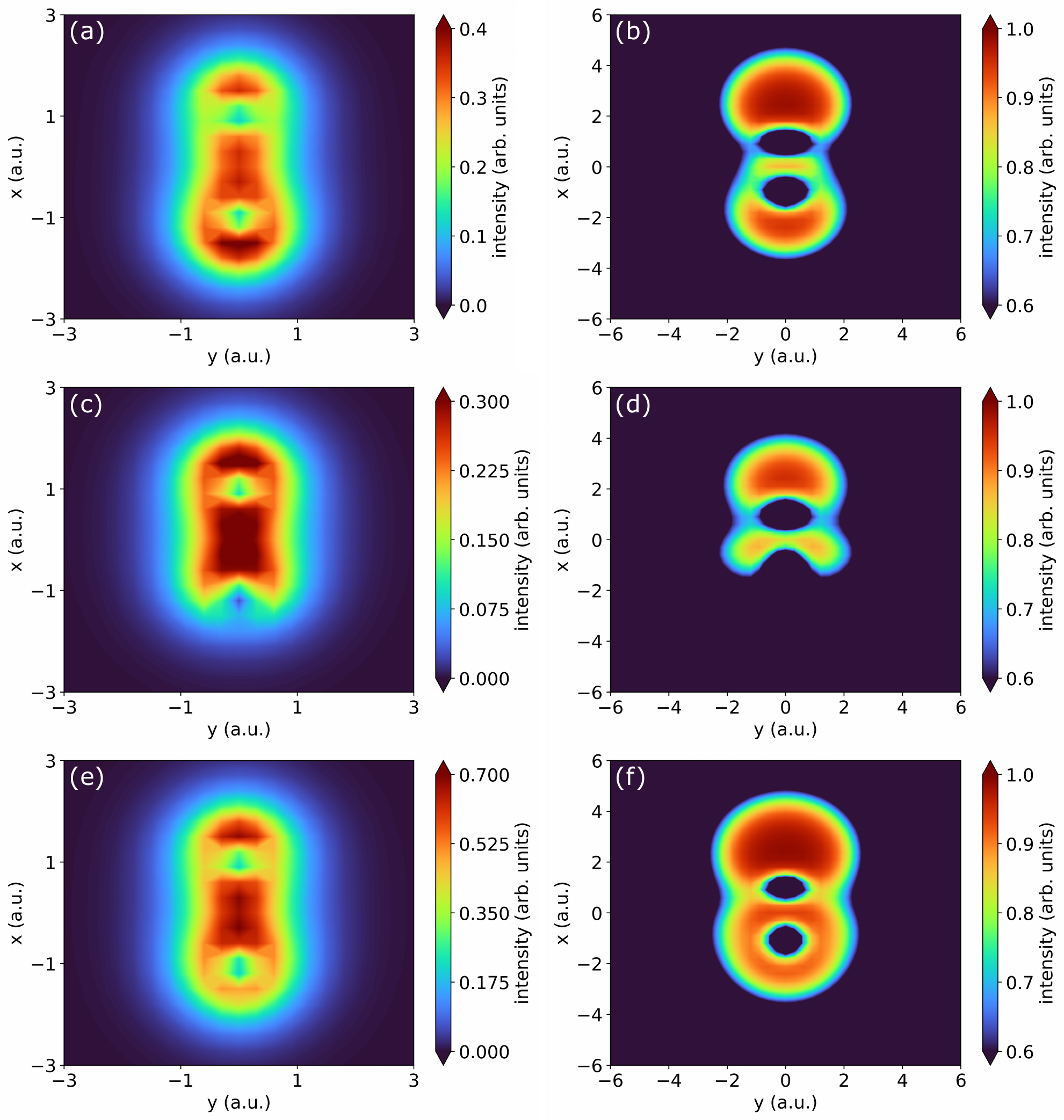}
\caption{\label{fig:iter11800_2} Cross sections (z=0) of (a) spin-up density, (c) spin-down density, (e) total density, (b) spin-up ELF, (d) spin-down ELF, and (f) total ELF along the xy-plane for the on-resonance, 5$\times$10$^{13}$ W/cm$^2$ case. These slices were taken at 5.7088 fs, which is near a zero of the laser field and a crossing between $2\sigma_u$ and $3\sigma_g$.}
\end{figure}

As there is more mixing with the $2\sigma_u$ the differences in the spin-up ELF decrease while the differences in spin-up density increase, see Figs. \ref{fig:iter11800_2}(a,b). Both the spin-down density and spin-down ELF start to go back to having more of a ''fish shape” as well shown in Figs. \ref{fig:iter11800_2}(c,d). The total ELF begins to have a decrease in the width in the bonding region, indicating an increase in triple bond character since the slice has more of the dumbbell shape typical for triple bonding. These characteristics can be seen in Fig. \ref{fig:iter11800_2}(f).

\begin{figure}[h!]
\includegraphics[width=0.9\columnwidth]{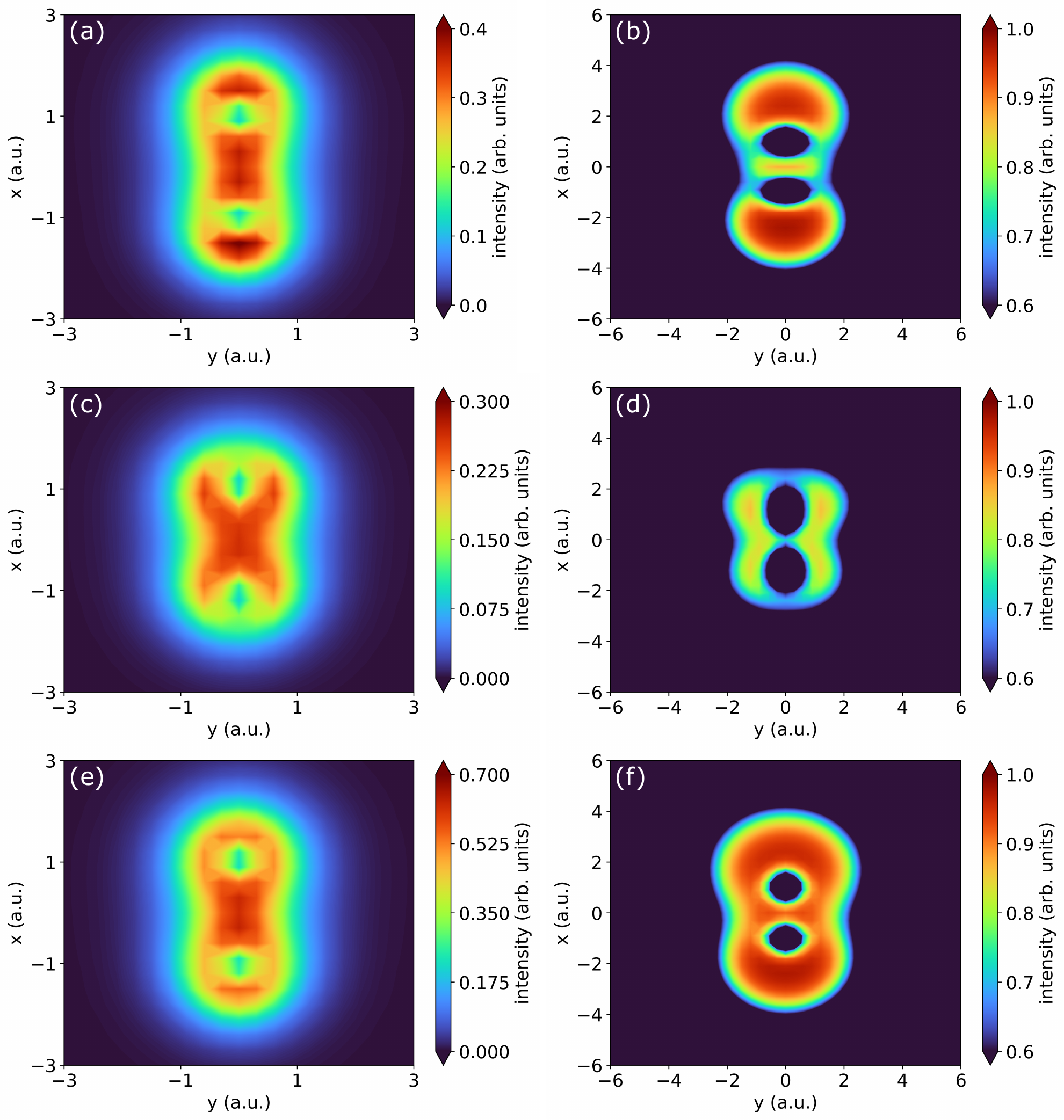}
\caption{\label{fig:iter15000_2} Cross sections (z=0) of (a) spin-up density, (c) spin-down density, (e) total density, (b) spin-up ELF, (d) spin-down ELF, and (f) total ELF along the xy-plane for the on-resonance, 5$\times$10$^{13}$ W/cm$^2$ case. These slices were taken at 7.2569 fs, which is near a minimum of the field during the period when $2\sigma_u$ is the filled orbital.}
\end{figure}

Once the system goes back to having the $2\sigma_u$ filled, there is an overall decrease in differences for both density and ELF, seen in Fig. \ref{fig:iter15000_2}. The bonding region in both goes largely unchanged during this time, showing that the changes are primarily coming from movement with the field. 

All of the same trends can be seen during the second periods of $2\sigma_u$ being filled (6-9 fs) and $3\sigma_g$ being filled (9+ fs). The only difference is that yet again the changes to the system decrease with decreasing field amplitude. The trends can also be seen for the higher peak intensity coupled simulation. The same visualizations for bonding still apply, but the changes occur more frequently due to the increase in Rabi flopping.

\begin{figure}[h!]
\includegraphics[width=0.8\columnwidth]{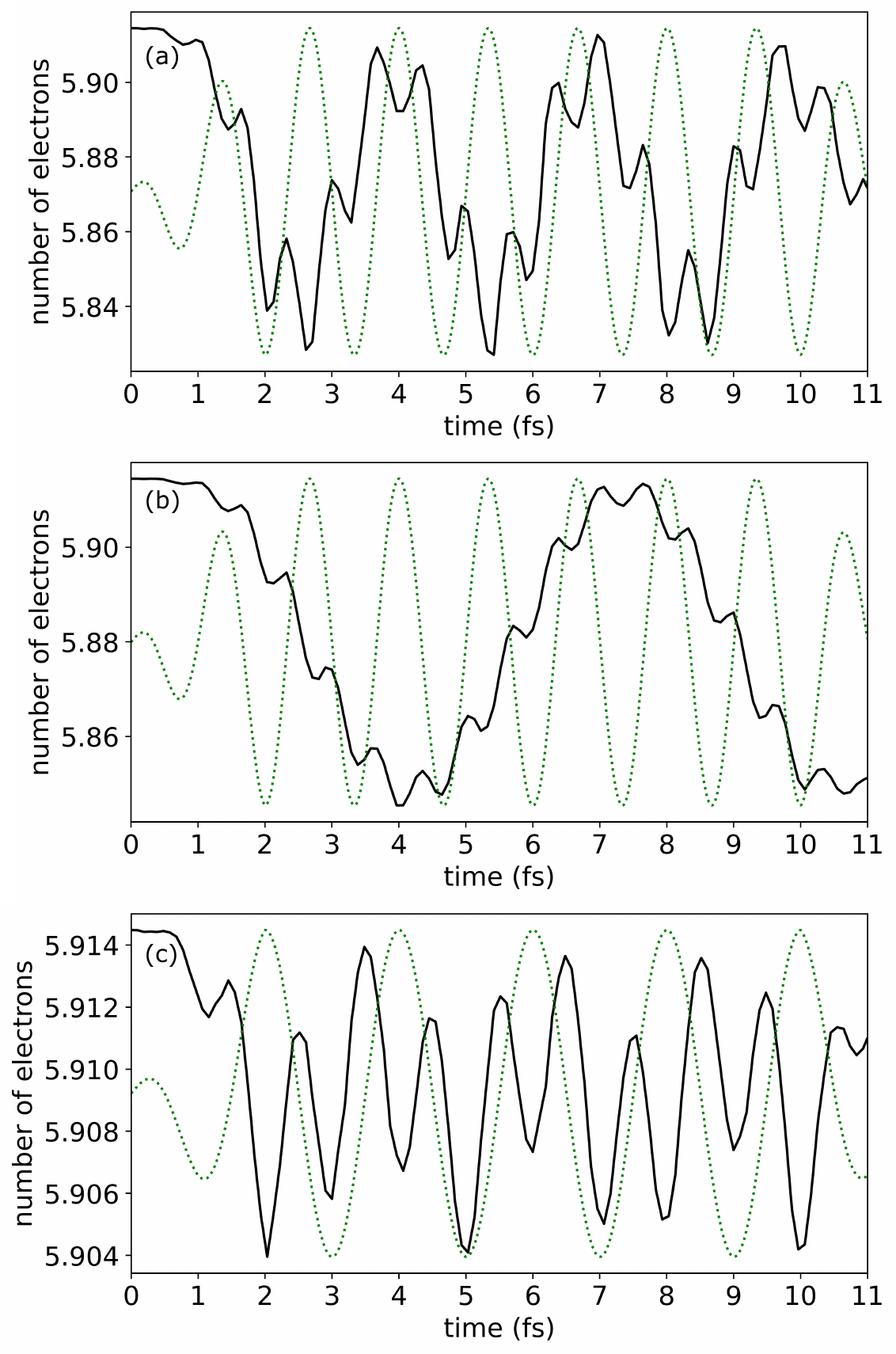}
\caption{\label{fig:integrateddens}
 Effect of localization on bond strength shown via integrated total density ($\rho_{\downarrow}(t) + \rho_{\uparrow}(t)$) in the bonding region ($x\in[-1.2, 1.2]$, $y\in[-40, 40]$, $z\in[-30, 30]$) plotted with the electric field in dotted green for (a) 400 nm, $2\times 10^{14} $ W/cm$^2$, (b) 400 nm, $5\times 10^{13} $ W/cm$^2$, and (c) 600 nm, $5\times 10^{13} $ W/cm$^2$.}
\end{figure}

The changes to the bonding shown in ELF match what can be seen with the integrated density in the bonding region shown in Fig. \ref{fig:integrateddens} for off- and on-resonance cases. Between approximately 2.5 and 6 fs, there is a prolonged decrease in the bonding region density for the on-resonance, 5$\times$10$^{13}$ W/cm$^2$ seen in Fig. \ref{fig:integrateddens}(b). This indicates a decrease in bond order which agrees with the change in shape of ELF from tube-shaped for $2\sigma_{u, \downarrow}$ being filled to dumbbell-shaped for $3\sigma_{g, \downarrow}$ being filled during that same time period and shown in Figs. \ref{fig:iter5400_2, fig:iter6200_2, fig:iter6800_2, fig:iter11800_2}. 

Figures \ref{fig:integrateddens}(a,b) also show this decrease in electron density in the bonding region being characteristic of the transition for both peak intensities because the shifting of the bond length density does not follow the field. Instead of following the laser field, there are extended periods of changes to the bond order, with additional temporal features from the laser field. For the non-resonant, 600 nm case the changes in the bond length density follow the field more regularly, without extended periods of decreased bond order seen in Fig. \ref{fig:integrateddens}(c).

\subsection{High Harmonic Generation}
The flopping between the two states seen in the time-dependent eigenvalues and projections (see Figs. \ref{fig:proj} and \ref{fig:energy}) is related to the Rabi frequency, which also causes Mollow sidebands in the harmonic spectra. The presence of these sidebands can act as another observable confirmation of the coupling since the spectra are dependent on the state of the system. The distance between the main harmonic and a sideband is the Rabi frequency as seen in Table \ref{tab:RabiFreqTable}.

\begin{table}[b]
\caption{\label{tab:RabiFreqTable} Rabi frequencies calculated and observed through the harmonic spectra. Calculated frequencies were found using $\Omega_{R} = \mu E_0$ where $\mu$ is the transition dipole and $E_0$ is the field amplitude.}
\begin{ruledtabular}
\begin{tabular}{cccc}
     $\lambda$ (nm)  & $I$ (W/cm$^2$)& $\Omega_{R,calc}$ (a.u.) & $\Omega_{R,obs}$ (a.u.) \\
     \hline
     400 & $5\times 10^{13}$ & 0.0592 &  0.0671 \\
     400 & $2 \times 10^{14}$ & 0.1184 &  0.1190 \\
     
\end{tabular}
\end{ruledtabular}
\end{table}

\begin{figure}[h!]
\includegraphics[width=1\columnwidth]{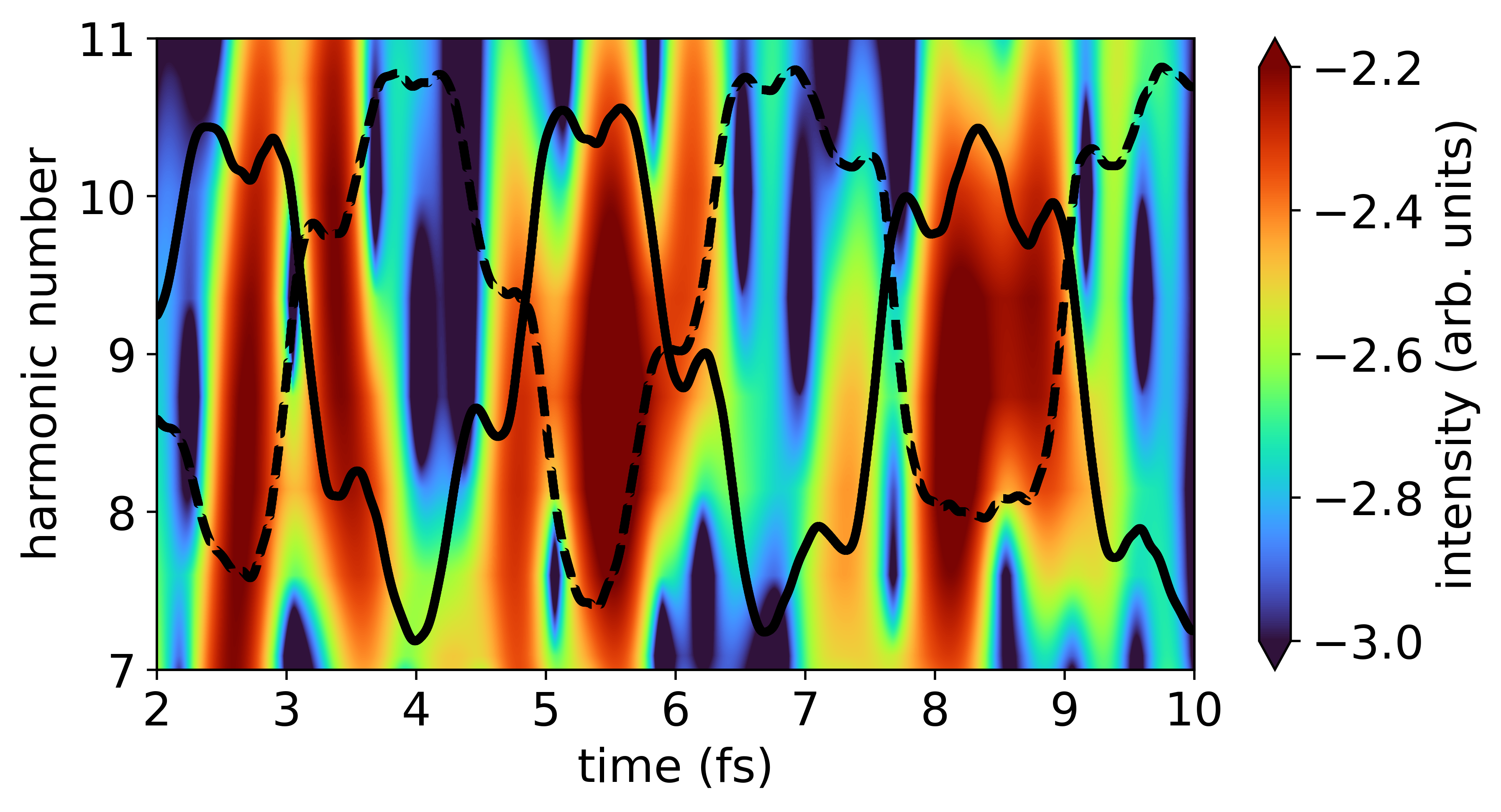}
\caption{\label{fig:400nm2e14Wavelet} Morse wavelet analysis for the on-resonance (400 nm), $2 \times 10^{14} $ W/cm$^2$ case with the time-dependent eigenvalues of $2\sigma_{u,\downarrow}$ (solid line) and $3\sigma_{g,\downarrow}$ (dashed line).}
\end{figure}

By taking the wavelet transform of the time-dependent dipole, it is possible to look at the time-dependent emission of the harmonics. The time-dependent emission of harmonics  allows  to correlate the temporal properties of the emission of harmonics to dynamics in the system. In present case a Morse wavelet transformation was applied to the total dipole.
Fig. \ref{fig:400nm2e14Wavelet} presents results for the time-dependent emission of the 9th harmonic and vicinity. It can be seen that 9th harmonic is emitted during periods when the $3\sigma_g$ orbital is filled. This shows that by controlling the states of the system with driving the resonance, it is possible to control emission of certain harmonic frequencies during the ultrashort pulses.

\subsection{Ionization and electronic flux}
Here we present results for total ionization related properties of the outgoing electronic flux. 

\begin{table}[b]
\caption{\label{tab:totalionization} The total ionization for each case where the total ionization is taken to be a percentage of an ensemble of molecules that would be ionized.}
\begin{ruledtabular}
\begin{tabular}{ccc}
     $\lambda$ (nm)  & $I$ (W/cm$^2$)& Total Ionization (\%) \\
    \hline
     600 & $5\times 10^{13}$ &  $8.07\times 10^{-5}$\\
     400 & $5\times 10^{13}$ &  $1.33 \times 10^{-4}$\\
     400 & $2 \times 10^{14}$ &  $5.57\times 10^{-2}$\\
\end{tabular}
\end{ruledtabular}
\end{table}

When the field is off-resonance with a peak intensity of $5\times 10^{13} $ W/cm$^2$, there is ionization on the order of $10^{-5}$. When the field is on-resonance for the same peak intensity, there is ionization on the order of $10^{-4}$. This increase by an order of magnitude in the ionization is related to the charge resonance enhanced ionization (CREI) \cite{Bandrauk_1995}. By tuning the field to a transition energy there is an increase in the ionization when compared to fields that are not tuned to a transition. The total ionization values can be seen in Table \ref{tab:totalionization}

Analysis of  the ionization can also make use of the outgoing electronic flux. Here we present results for both off- and on-resonance cases. The outgoing electronic flux is calculated according to the equation, 
\begin{equation}
    f_{flux} = - \frac{d}{dt} \iiint \rho(\textbf{x},\textbf{y},\textbf{z},t) dx dy dz, 
    \label{eqn:ionrate}
\end{equation}
where $\rho(\textbf{x},\textbf{y},\textbf{z},t)$ is the total density. The limits in $\hat{y}$ and $\hat{z}$ encompass the entirety of the simulation box in those directions, $\pm $40 a.u. and $\pm$ 30 a.u. respectively. Integration in the $\hat{x}$-direction was limited to $\pm$ 6.9 a.u. to extend beyond the quiver radius, where the outgoing wavepacket can be considered ionized, for all calculations. To save computational resources, the density was outputted after every 200th iteration of the calculation or 0.0968 fs.

\begin{figure}[h!]
    \includegraphics[width=1\columnwidth]{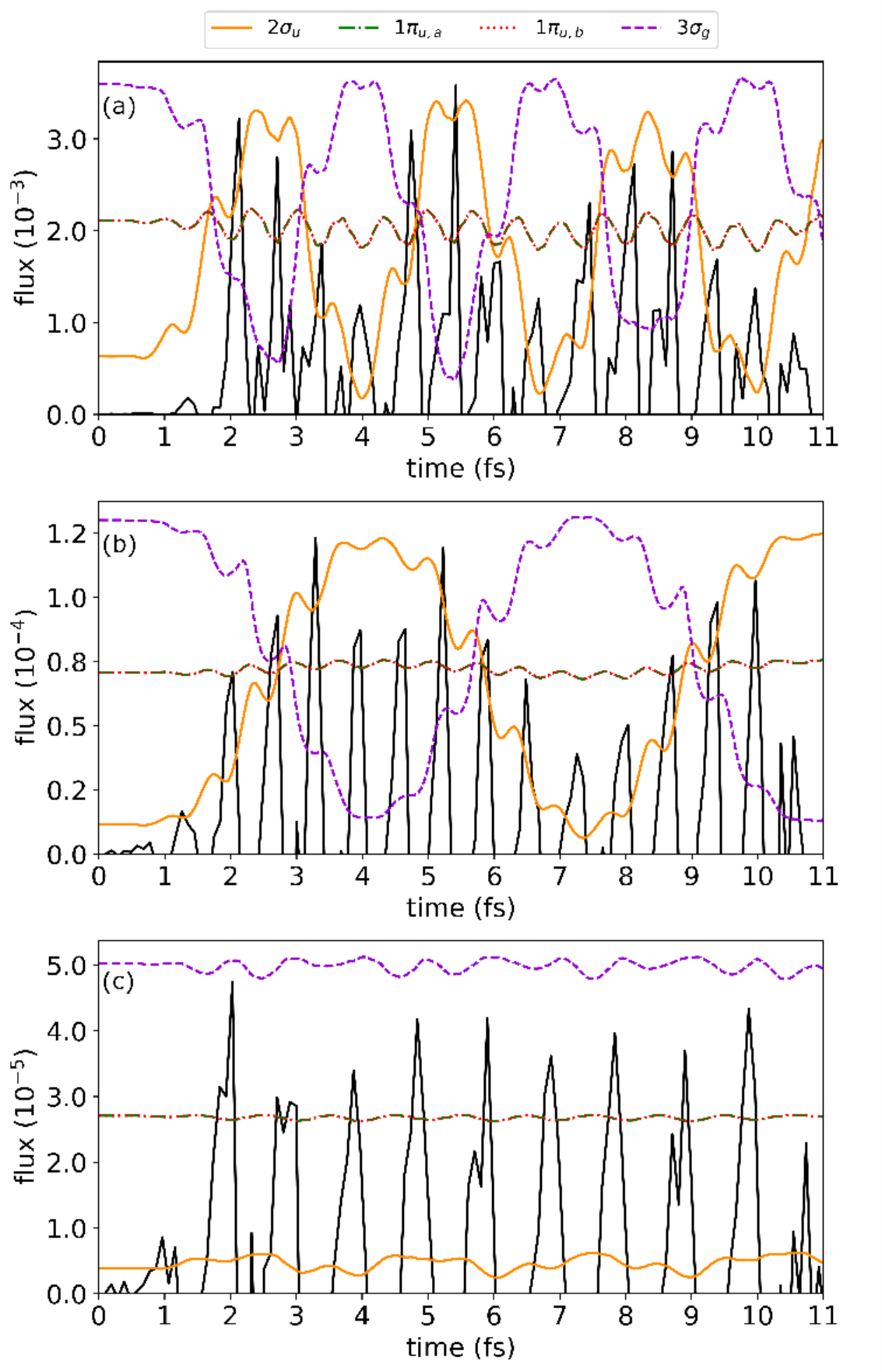}
    \caption{\label{fig:flux} Total electronic flux calculated using Eq. \ref{eqn:ionrate} with the spin-down, time-dependent eigenvalues for (a) 400 nm, $2\times 10^{14} $ W/cm$^2$, (b) 400 nm, $5\times 10^{13} $ W/cm$^2$, and (c) 600 nm, $5\times 10^{13} $ W/cm$^2$.}
\end{figure}

The bursts of electronic flux occur at the extrema of the pulse, as expected. Figures \ref{fig:flux}(a) and \ref{fig:flux}(b) show that the outgoing electronic flux is higher when the $3\sigma_g$ is the filled orbital. This is another indication that CREI is occurring. For the off-resonance case, the electronic flux is consistent throughout as seen in Fig. \ref{fig:flux}(c)

\begin{figure}[h!]
\includegraphics[width=1\columnwidth]{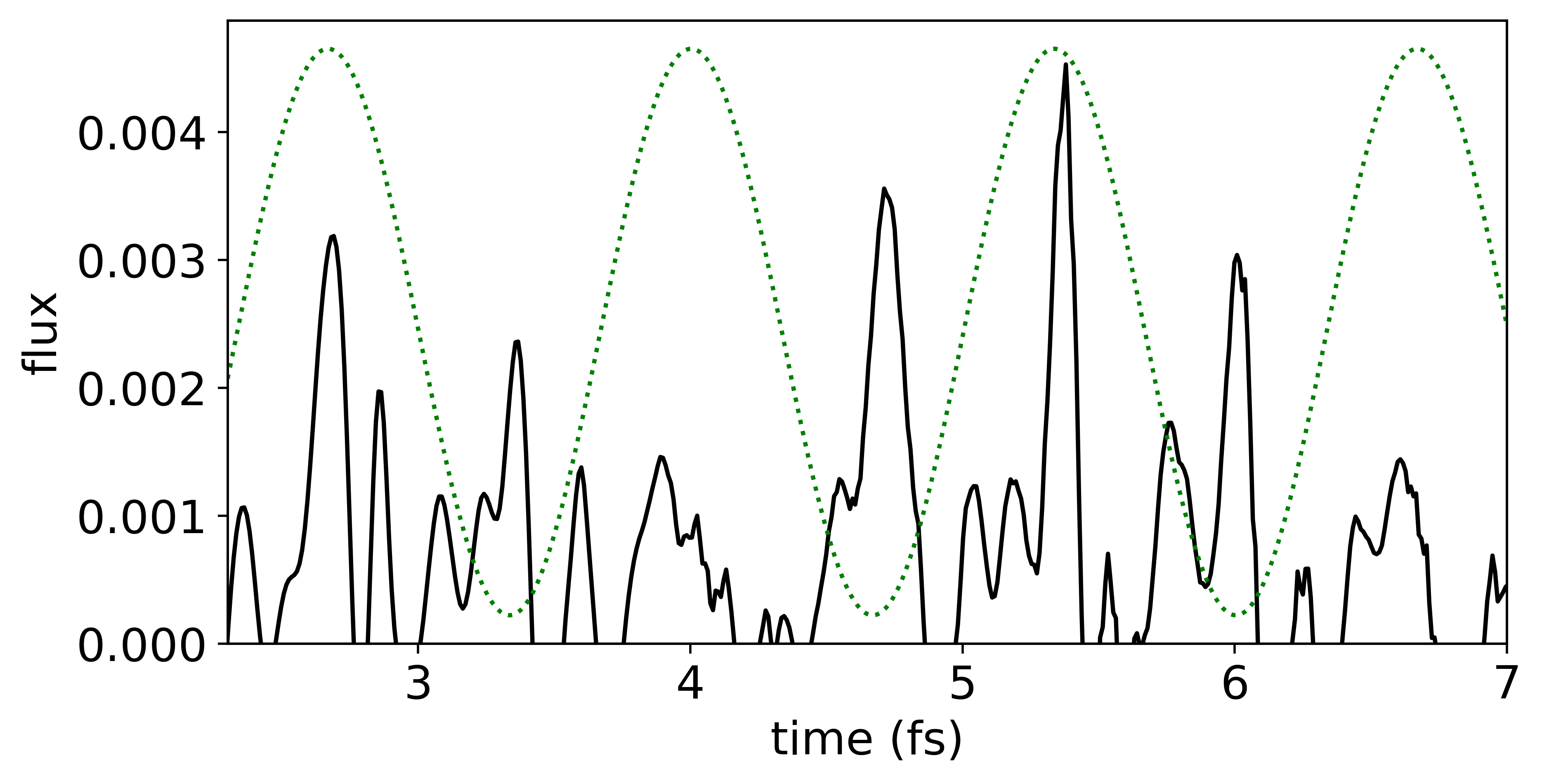}
\caption{Electronic flux calculated using Eq. \ref{eqn:ionrate} for the on-resonance 400 nm, $2 \times 10^{14} $ W/cm$^2$ case using the total density outputted every 20 iterations (0.00968 fs). The field is plotted with the dotted green line.}
\label{fig:ionrate20iter}
\end{figure}

Multiple bursts of ionization were expected, as it has alo been previously seen in case of H$_2^+$ \cite{Takemoto_2010}. Figure \ref{fig:flux}(a) shows signs of multiple bursts of ionization with wide peaks that have an almost trapezoidal shape with two local maxima instead of only having one local maximum. In order to better see the potential instances of multiple burst ionization, a finer output was used where the density was outputted after every 20th iteration or 0.00968 fs. Figure \ref{fig:ionrate20iter} shows that this finer output of densities does resolve the peaks. This is an indication that the multiple bursts are present whoever occur on faster timescales than those seen with previously H$_2^+$.

\subsection{Time Dependent Average Local Ionization Energy}

In order to better understand how the changes to the molecular system correspond to suppressed and enhanced ionization shown through the electronic flux in previous sub-section, the average local ionization energy was calculated. The average local ionization energy $\bar{I}(\textbf{r})$ is equal to the ratio of the sum of single-particle energy densities for all occupied orbitals and the total density, 
\begin{equation}
    \label{eq:ALIE}
    \bar{I}(\textbf{\textbf{r}}) = \frac{\sum_{i} \rho_{i}(\textbf{\textbf{r}}) |\varepsilon_{i}|}{\rho(\textbf{r})},
\end{equation}
where $\rho_{i}(r)$ is the density of the orbital $\phi_{i}(r)$ with energy $\varepsilon_{i}$ and the total density $\rho(r)$. This representation is valid under the assumption that the ionization potential of orbitals can be approximated by the orbital energy. Originally formulated in the Hartree-Fock framework \cite{Politzer_1991} it was shown that a DFT formulation also has physical meaning \cite{Politzer_1998}.

Let us note that the average local ionization energy (ALIE) can be used as a guide for reactivity in molecules. For example areas where $\bar{I}(r)$ is relatively low are indicative of more loosely bound electrons and these areas are expected to be the most reactive. Applications of ALIE have been used to evaluate differences between saturated and unsaturated hydrocarbon rings \cite{Murray_1990} and to identify the more energetic electrons in a molecule \cite{Bulat_2021}.

The TDALIE was used to analyze N$_2^+$ to get a better understanding of the changes to the electronic flux or ionization rate. The time-dependent case is obtained by taking static case and  calculating with corresponding time dependent counterparts,  
\begin{equation}
    \label{eq:TDALIE}
    I(\textbf{r},t) = \frac{\sum_{i} \rho_{i}(\textbf{r},t) |\varepsilon_{i}(t)|}{\rho(\textbf{r},t)}.
\end{equation}

\begin{figure}[h!]
    \includegraphics[width=1\columnwidth]{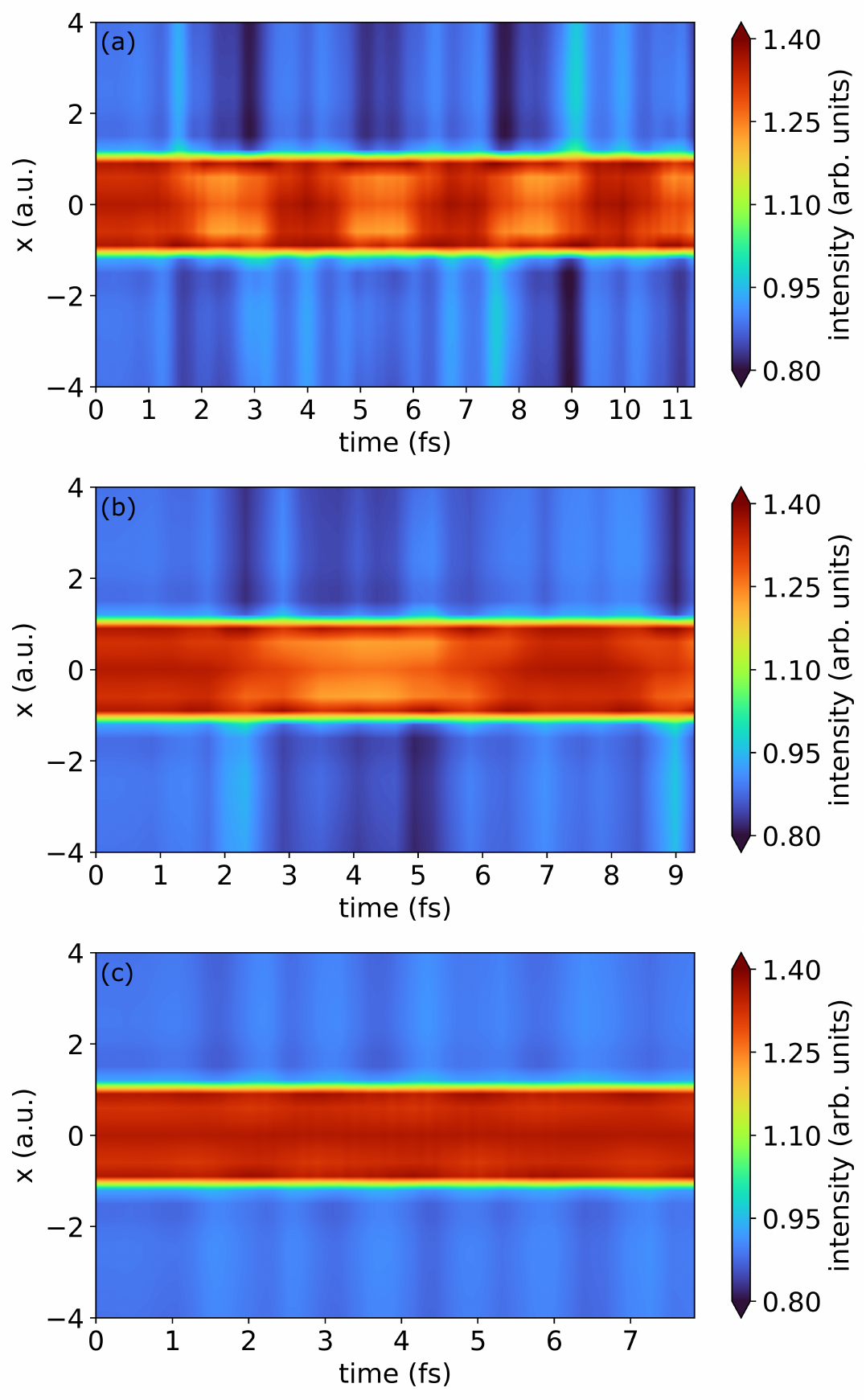}
    \caption{\label{fig:TDALIE} Time-dependent average local ionization energy (TDALIE) calculated using Equation \ref{eq:TDALIE} and plotted as slices along the molecular axis (y=0, z=0) for (a) 400 nm, $2\times 10^{14} $ W/cm$^2$, (b) 400 nm, $5\times 10^{13} $ W/cm$^2$, and (c) 600 nm, $5\times 10^{13} $ W/cm$^2$.}    
\end{figure}

For the non-resonant, 600 nm  calculations there are only small modulations in the TDALIE shown in Fig. \ref{fig:TDALIE}(c). These changes largely correspond to oscillations with the field which supports the ionization rate being consistent throughout the duration of the pulse. For both on-resonant, 400 nm calculations there are distinct periods of the TDALIE decreasing in the bonding region shown in Figs. \ref{fig:TDALIE}(b) and \ref{fig:TDALIE}(a). The periods of TDALIE lowering do not correspond to minima and maxima in the field, indicating it is at least partially due to the resonance and consequence of the electron localization. 

\begin{figure}
 \includegraphics[width=1\columnwidth]{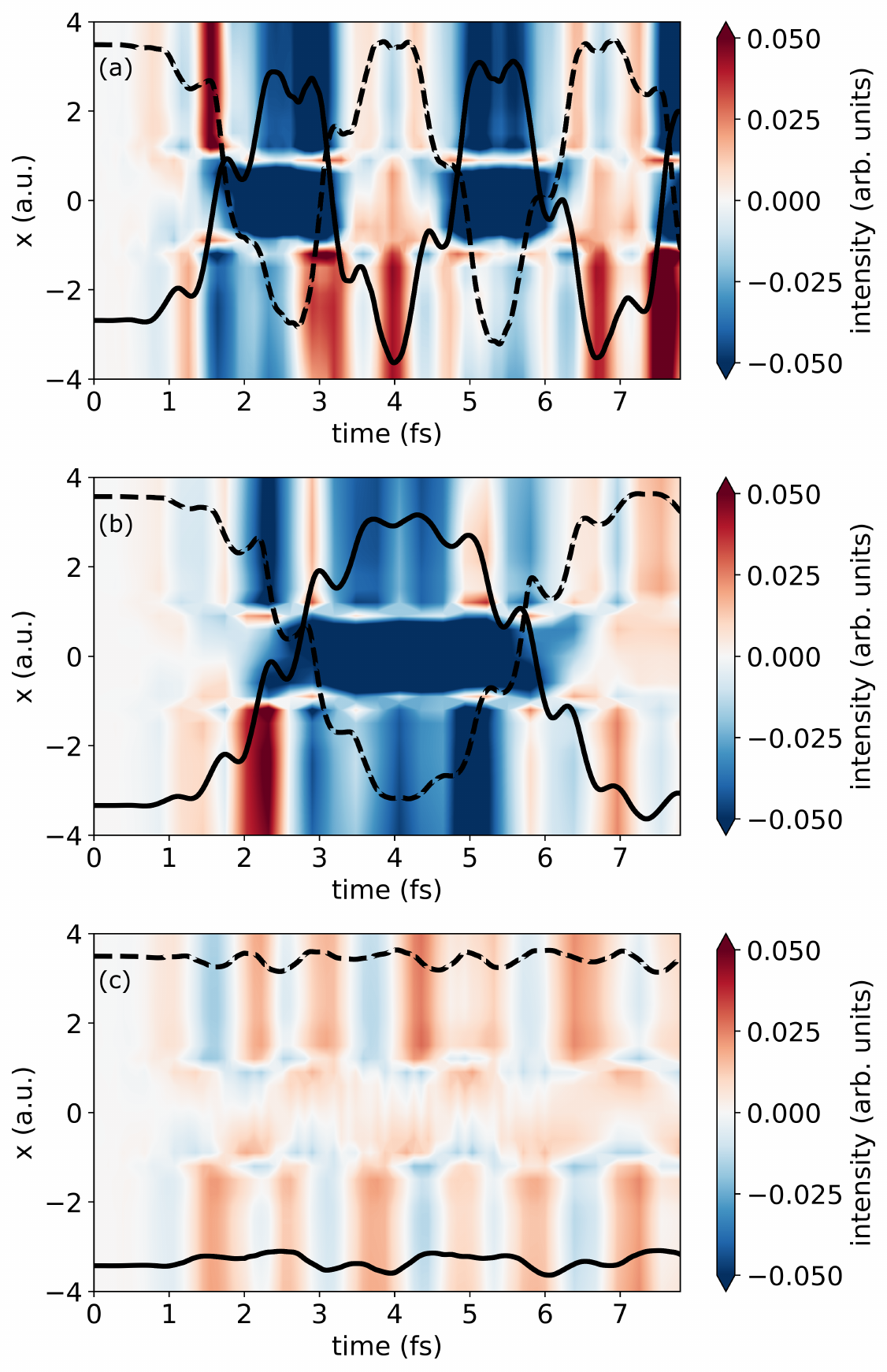}
\caption{\label{fig:TDALIEdiff} Time-dependent average local ionization energy (TDALIE) difference taken to be $\bar{I}(\textbf{r},t) - I(\textbf{r},t=0)$ and plotted as slices along the molecular axis (y=0, z=0) for (a) 400 nm, $2\times 10^{14} $ W/cm$^2$, (b) 400 nm, $5\times 10^{13} $ W/cm$^2$, and (c) 600 nm, $5\times 10^{13} $ W/cm$^2$. The time-dependent eigenvalues for the $2\sigma_u$ (solid line) and $3\sigma_g$ (dashed line) orbitals are plotted as well.}
\end{figure}

In order to get a clearer picture of the relationship between the ALIE and ionization rate, the TDALIE difference was plotted with the time-dependent eigenvalues for the $2\sigma_u$ (dashed line) and $3\sigma_g$ (dotted line) orbitals. Looking at Fig, \ref{fig:TDALIEdiff} it can be seen that there is a decrease (blue) in the ALIE during periods where the population changes from $2\sigma_u$ to $3\sigma_g$ for both peak intensities investigated.
The periods of decreased TDALIE allow for the emitted wavepacket to have higher kinetic energy. The increase in kinetic energy increases the velocity of the outgoing wavepacket which can partially be the reason there is an increase in ionization rate during the $3\sigma_g$ filled periods seen in Fig. \ref{fig:flux}.

\section{Summary}
We have presented results that indicate nonadiabatic dynamics that can be used for system control. By tuning to a transition within N$_2^+$ it was shown that there are effects on high harmonic generation, ionization, and bonding due to the localization of the electron wavepacket. We conclude that there are observable characteristic features of the resonance, indicating the state of the system and these observables can be related to theoretical methods. The changes due to dynamic electron localization (for 400nm) are contrasted with the typical dynamics driven by the ultrashort intense 600 nm laser pulse. TDELF and TDALIE as well as time dependent density and flux are used to visualize the interaction of the pulse with the electrons in the molecule in molecular region and rescattering region. 

\section{Acknowledgements}
A. J. acknowledge support from National Science Foundation PHY-2317149 and PHY-2110628 awards. 
Funding during the
final months of the project was provided by the Chemical Sciences,
Geosciences, and Biosciences Division, Office of Basic Energy Sciences,
Office of Science, U.S. Department of Energy, Grant no. DE-FG02-86ER13491.

Simulations were performed on the Alpine high-performance computing resource jointly funded by the University of Colorado Boulder, the University of Colorado Anschutz, and Colorado State University.

\bibliography{apssamp}
% Produces the bibliography via BibTeX.

\end{document}